\documentclass[11pt]{article} 
\usepackage{fullpage}
\usepackage{amsmath,amsfonts,amssymb}
\usepackage{euscript}
\usepackage[colorlinks=true,bookmarks=false,citecolor=blue,urlcolor=blue]{hyperref} 
\numberwithin{equation}{section}
\usepackage{subeqnarray}
\usepackage[utf8]{inputenc}
\usepackage[english]{babel}
\usepackage{csquotes}
\usepackage{cite} % combines consecutive citation references into ranges.

\newcommand{\LL}{\mathcal{L}}
\newcommand{\del}{\partial}
\newcommand{{\bb}}{\mathfrak{b}}
\newcommand{{\cc}}{\mathfrak{c}}

\DeclareMathOperator{\sgn}{sgn}

\usepackage{scalerel,stackengine}
\stackMath
\newcommand\reallywidehat[1]{%
\savestack{\tmpbox}{\stretchto{%
  \scaleto{%
    \scalerel*[\widthof{\ensuremath{#1}}]{\kern-.6pt\bigwedge\kern-.6pt}%
    {\rule[-\textheight/2]{1ex}{\textheight}}
  }{\textheight}% 
}{0.5ex}}%
\stackon[1pt]{#1}{\tmpbox}%
}

\begin{document}

\title{Diffeomorphism covariance and the quantum Schwarzschild interior}
\author{I. W. Bornhoeft${}^1$, R. G. Dias${}^2$, and J. S. Engle${}^3$ \\
Department of Physics, Florida Atlantic University, \\ 777 Glades Road, Boca Raton, FL 33431, USA
}

\maketitle

\footnotetext[1]{ibornhoeft2018@fau.edu}
\footnotetext[2]{rdias2019@fau.edu}
\footnotetext[3]{jonathan.engle@fau.edu}

\begin{abstract}
We introduce a notion of residual diffeomorphism covariance in quantum Kantowski-Sachs (KS), describing the interior of a Schwarzschild black hole. We solve for the family of Hamiltonian constraint operators satisfying the associated covariance condition, as well as parity invariance, preservation of the Bohr Hilbert space of Loop Quantum KS and a correct (na\"ive) classical limit.
We further explore imposing minimality of the number of terms, and compare the solution with other Hamiltonian constraints proposed for Loop Quantum KS in the literature. 
In particular, we discuss a lapse recently commonly chosen due to the resulting decoupling of evolution of the two degrees of freedom and exact solubility of the model. We show that such a lapse choice can indeed be quantized as an operator densely defined on the Bohr Hilbert space, and that any such operator must include an infinite number of shift operators. 
\end{abstract}

\section{Introduction}

A central epistemic value in science is that of \textit{simplicity} --- that a theory be derived uniquely from as few principles as possible. It follows that it is important to eliminate, inasmuch as possible, ambiguities present in a theory through physical principles, in particular when observational data is scarce.

General Relativity is based on background independence, which is equivalent to covariance under diffeomorphisms \cite{Rovelli:2004tv}. Guided by this principle, Loop Quantum Gravity (LQG) is a non-perturbative approach to a quantum theory of gravity \cite{Ashtekar:2004eh, Rovelli:2004tv, Thiemann:2007pyv, Gambini:2011zz,  Rovelli:2014ssa,  Ashtekar:2017yom}, and Loop Quantum Cosmology (LQC) models arise from applying quantization techniques analogous to LQG to symmetry-reduced gravitational models \cite{Bojowald:2008zzb,Ashtekar:2011ni,Agullo:2016tjh}. In order to ensure that a given LQC model faithfully reflects the diffeomorphism covariance of full loop quantum gravity, it is important for this model to also be diffeomorphism covariant in some sense, a requirement which can also serve to reduce ambiguities in its construction.  
Related prior work along these lines include:
\begin{itemize}
    \item Lewandowski, Okolow, Sahlmann, and Thiemann \cite{Lewandowski:2005jk} proved that the requirement of invariance under spatial diffeomorphisms, or more precisely, the unitary implementation of the action of the diffeomorphism group, establishes the uniqueness of the kinematics of LQG.
    \item For LQC, Ashtekar and Campiglia \cite{Ashtekar:2012cm} showed that, in the case of the Bianchi I model, a unique kinematical representation is achieved through invariance under canonical, and thus volume preserving, \textit{residual diffeomorphisms}, i.e., diffeomorphisms not frozen by the gauge fixing required by symmetry reduction.
    \item The works \cite{Engle:2016hei,Engle:2016zac} extended the result to single out the standard kinematical Hilbert space of the homogeneous isotropic case, by requiring invariance under also \textit{non-canonical} residual diffeomorphisms.
    \item The works \cite{Engle:2018zbe, Engle:2019zfp} demonstrated, for the cases of homogeneous isotropic LQC and Bianchi I models, that a family also of \textit{dynamics} can be derived from residual diffeomorphism covariance, and, 
    if desired, uniqueness can be achieved by requiring \textit{minimality} 
    --- a form of Occam's razor, requiring the Hamiltonian have a minimal number of terms, i.e., a minimal number of shift operators --- in addition to a further assumption of planar loops for the Bianchi I case.
\end{itemize}

In this work, we investigate the choice of dynamics for the loop quantization of the Schwarzschild black hole interior, described by the Kantowski-Sachs (KS) framework. There is a wide literature discussing different proposals for such a choice (for instance, \cite{Modesto:2005zm, Ashtekar:2005qt, Campiglia:2007pb, Chiou:2008eg, Joe:2014tca, Corichi:2015xia, Cortez:2017alh, Ashtekar:2018cay, Bodendorfer:2019cyv, Sartini:2020ycs, Ashtekar:2023cod}).
Here, instead of quantizing the Hamiltonian directly, we narrow the possibilities by imposing physically motivated properties, namely the following:
\begin{itemize}
    \item 
Covariance under residual diffeomorphisms. Looking at how the phase space variables flow under the action of the residual diffeomorphisms (Section \ref{Residual Diffeomorphisms}), we formulate a condition for the covariance of the Hamiltonian, which we quantize, establishing a condition of \textit{quantum covariance} under such diffeomorphisms. The residual diffeomorphisms are non-canonical, so that this requires novel methods (Section \ref{Covariance Equation}).
    \item Covariance under discrete residual automorphisms of the $SU(2)$ principal fiber bundle (Section \ref{Discrete Symmetries}). 
    \item The correct (na\"ive) classical limit (Section \ref{Classical asymptotic behavior}).
\end{itemize}
In addition to these basic physical criteria, we also consider the consequences of the additional criterion of \textit{minimality}, that the quantum Hamiltonian constraint contain a minimal number of shift operators (Section \ref{Minimality}).

The na\"ive classical limit, which has been used in all LQC and 
loop quantum KS literature up until now, corresponds to 
$\hbar \rightarrow 0$, or the limit in which the Planck length $\ell_P:=\sqrt{G\hbar}$ goes to zero, or, equivalently, the limit under which the length
of curves regularizing curvature and connection factors in the 
Hamiltonian constraint go to zero, 
which in section \ref{low curvature} we show is equivalent to the eigenvalues
of extrinsic curvature going to zero.
This is the definition of classical limit used in this paper, 
as the focus of this paper is not to develop a new one. 
However, this definition of classical limit is limited, 
because the truly relevant criterion for the classical 
regime is that \textit{four-dimensional} curvature scalars
should go to zero, which can happen even if the eigenvalues 
of extrinsic curvature do not. 
Indeed, in KS, this happens at the horizon, 
where the latter diverge, while the former remain small compared 
to the Planck scale. 
This is in fact the regime in which the model of Ashtekar, Olmedo, and Singh (AOS) \cite{Ashtekar:2018cay,Ashtekar:2023cod}, as well as the earlier models \cite{Corichi:2015xia,Olmedo:2017lvt},
perform better than all other models, and this is remarked upon 
in section \ref{aossect}.

In Section \ref{Choice of Lapse} we present a discussion of a choice of lapse used in the literature that greatly simplifies the classical and (with further assumptions) effective equations, rendering them analytically solvable. 
In particular, we prove that the quantum Hamiltonian operator resulting from such a choice can be densely defined on the usual Bohr Hilbert space motivated by by loop quantum gravity, and only with an infinite number of shift operators. 

For completeness, and to fix notation, we start with a background review of the KS framework and its loop quantum kinematics (Section \ref{Background}).
To finish, we compare our conclusions with other proposals in literature (Section \ref{Comparison Literature}).

\section{Background}\label{Background}
\subsection{Kantowski-Sachs in Ashtekar-Barbero variables}

The interior region of a Schwarzschild black hole can be foliated in homogeneous 3-manifolds, of topology $\mathbb{R}\times S^2$, invariant under the Kantowski-Sachs group $\mathbb{R}\times SO(3)$. 
We introduce standard coordinates $(\theta,\phi)$ on the $S^2$ factor, and a coordinate $x$ on the $\mathbb{R}$ factor, as well as a fiducial cell of coordinate length $L_o$ in the non-compact $x$ direction as an infrared cutoff, to prevent integrations from diverging. Physical results are required to be indepedent of this parameter. 

Geometry is characterized by a symmetry-reduced phase space described by two conjugate pairs of variables $(b,p_b)$ and $(c,p_c)$, with 
%symplectic structure
%\begin{align*}
%    \Omega = \frac{1}{2\gamma G}\left[2db\wedge dp_b + dc\wedge dp_c\right], 
%\end{align*}
%yielding the 
Poisson brackets 
  \begin{align}\label{Symplectic structure}
        \{b,p_b\} = G\gamma\hspace{0.5cm} &, \hspace{0.5cm} \{c,p_c\} = 2G\gamma.
    \end{align}
In terms of these, the Ashtekar-Barbero connection and densitized triad are given by
    \begin{align}\label{KS Ashtekar variables}
    A_a^1 = -b\sin\theta\partial_a\phi\hspace{0.5cm} &, \hspace{0.5cm} E^a_1 = -\frac{p_b}{L_0}\phi^a, \nonumber \\
    A_a^2 = b\partial_a\theta \hspace{0.5cm} &, \hspace{0.5cm} E^a_2 = \frac{p_b}{L_0}\sin\theta \theta^a, \\
    A_a^3 = \frac{c}{L_0}\partial_a x + \cos\theta\partial_a\phi \hspace{0.5cm} &, \hspace{0.5cm} E^a_3 = p_c\sin\theta x^a \nonumber
\end{align}
where $\phi^a, \theta^a, x^a$ denote the coordinate vector fields. 
The corresponding homogeneous spacetime metric is given by    
\begin{align}\label{KS metric}
        ds^2=-N^2 d\tau^2 + \frac{p_b^2}{|p_c|L_0^2}dx^2 + |p_c|d\Omega^2 ,
\end{align}
 which can be identified with the Schwarzschild interior metric
\begin{align}
   ds^2 = -\left(\frac{2m}{\tau}-1\right)^{-1}d\tau^2+\left(\frac{2m}{\tau}-1\right)dx^2+\tau^2d\Omega^2, \nonumber
\end{align}
for $\tau < 2m$, by choosing the lapse and consequent evolution of the momenta to be
    \begin{align}\label{Schwarszchild relation}
    |p_c| = \tau^2, \hspace{1cm} p_b^2=L_0^2\left(\frac{2m}{\tau}-1\right)\tau^2, \hspace{1cm} N^2=\left(\frac{2m}{\tau}-1\right)^{-1}.
    \end{align}
    
From \eqref{KS Ashtekar variables}, one can calculate the Hamiltonian constraint to be
\begin{align}\label{Hcl}
    H_{c\ell}[N] &   = -\frac{N}{2G\gamma^2}\frac{b\sgn p_c}{\sqrt{|p_c|}}\left(p_b\left(b + \frac{\gamma^2}{b}\right) + 2cp_c\right), 
\end{align}
for an arbitrary lapse $N$. 
Letting $V = 4\pi|p_b|\sqrt{|p_c|}$ denote the physical volume of the fiducial cell, we choose a family of lapses of the form
\begin{align}
    \label{lapse volume} N = V^n =  \left(4\pi|p_b|\sqrt{|p_c|}\right)^n,
\end{align}
for $n>-3$. This covers the cases of proper time ($n=0$, as in \cite{Ashtekar:2005qt, Joe:2014tca}) and harmonic time gauge ($n=1$, as in \cite{Chiou:2008eg}), among others --- for instance, the case $n=-1$ appears when considering unimodular gravity \cite{Sartini:2020ycs}. From now on we will assume this choice of lapse and simply represent $H_{c\ell}[N]$ as $H_{c\ell}$. There is a choice of lapse prominent in the literature that does not fall into this family; we discuss this choice in section \ref{Choice of Lapse}. The restriction $n>-3$ will be needed in section \ref{quantum covariance section}. The classical Hamiltonian constraint (\ref{Hcl}) then becomes
\begin{align}\label{Hcl NV}
    H_{c\ell}& = -\frac{(4\pi)^n|p_b|^n|p_c|^{\frac{(n-1)}{2}}b\sgn p_c}{2G\gamma^2}\left(p_b\left(b + \frac{\gamma^2}{b}\right) + 2cp_c\right) \nonumber \\
    & = - \frac{V^{n+1}}{8\pi G\gamma^2}\sgn p_b\left(\frac{b^2+\gamma^2}{p_c} + \frac{2bc}{p_b}\right).
\end{align}

\subsection{Quantum Kinematics}\label{Quantum Kinematics}

The basic configuration variables with direct quantum analogues in loop quantum gravity and loop quantizations of symmetry reduced models are always some class of holonomies $h_e[A]$. For the Kantowski-Sachs framework, one consider holonomies along curves parallel to the $x$ axis, along $x=$constant curves that are geodesic with respect to the 2-sphere metric: 
\begin{align}\label{set holonomies}
    h_x[A] &= \exp\left(-i\frac{\tau c}{2}\sigma_3\right) = \cos\left(\frac{\tau c}{2}\right)\mathbb{I} + 2\sin\left(\frac{\tau c}{2}\right)\tau_3 \nonumber \\
    h_\theta[A] &= \exp\left(-i\frac{\mu b}{2} \sigma_2\right) = \cos\left(\frac{\mu b}{2}\right)\mathbb{I} + 2\sin\left(\frac{\mu b}{2}\right)\tau_2 \\
    h_\phi[A]|_{\theta = 90°} &= \exp\left(i\frac{\mu b}{2}\sigma_1\right) = \cos\left(\frac{\mu b}{2}\right)\mathbb{I} - 2\sin \left(\frac{\mu b}{2}\right)\tau_1. \nonumber
\end{align}
The matrix elements of these holonomies generate the algebra of \textit{almost periodic functions}, composed by elements of the form
\begin{align}
\label{apfunction}
        f(b,c) = \sum_{j=1}^N f_je^{\frac{i}{2}(\mu_j b + \tau_jc)},
\end{align}
with $N$ possibly infinite, $f_j\in\mathbb{C}$ and $\mu_j,\tau_j\in\mathbb{R}$. 
The space of such functions, endowed with, and normalizable with respect to, the inner product
\begin{align*}
    \left\langle e^{\frac{i}{2}(\mu_j b + \tau_jc)}\right|\!\left.e^{\frac{i}{2}(\mu_k b + \tau_kc)}\right\rangle \! := \! \lim_{L\rightarrow \infty} \frac{1}{(2L)^2}\!
    \int_{(-L,L)^2} \hspace{-2.5em} \overline{e^{\frac{i}{2}(\mu_j b + \tau_jc)}} e^{\frac{i}{2}(\mu_k b + \tau_kc)} \text{d}b \text{d}c =  \delta_{\mu_j}^{\mu_k}\delta_{\tau_j}^{\tau_k},
\end{align*}
is called the \textit{Bohr Hilbert space}, denoted $\mathcal{H}_{\text{Bohr}}$, the 
space of kinematical states for the quantum theory.
The momenta are quantized as
\begin{align}\label{KS momentum operator}
    \hat{p}_b = -i\gamma\ell_{P}^2\frac{\del}{\del b},
    \qquad
    \hat{p}_c = -2i\gamma\ell_{P}^2\frac{\del}{\del c}, 
\end{align}
so that each associated normalized simultaneous eigenstate $|p_b,p_c\rangle$ has wavefunction 
$\psi_{p_b,p_c}(b,c) = e^{\frac{i}{\gamma \ell_P^2}\left(p_b b+\frac{p_c c}{2}\right)}$. \eqref{apfunction} can then also be written
\begin{align}
        f(b,c) %= \sum_{j=1}^N f_je^{\frac{i}{\gamma \ell_{P}^2}\left(p_b^jb + \frac{1}{2}p_c^jc\right)}   
        = \sum_{j=1}^N f_j|p_b^j,p_c^j\rangle.
\end{align}
Complex exponentials of $b$ and $c$ then act as shift operators by
\begin{align*}
  e^{i\lambda b}|p_b^j,p_c^j\rangle = |p_b^j+\gamma\ell_P^2\lambda,p_c^j\rangle \hspace{1cm} \textrm{and}\hspace{1cm} e^{i\eta c}|p_b^j,p_c^j\rangle = |p_b^j,p_c^j+2\gamma\ell_P^2\eta\rangle.  
\end{align*}

\section{Residual Diffeomorphisms}\label{Residual Diffeomorphisms}

The kinematical symmetry group of the Ashtekar-Barbero formulation of gravity is the group $\text{Aut}$ of automorphisms of the $SU(2)$ principle bundle, isomorphic to the semi-direct product of diffeomorphisms of the spatial slice and $SU(2)$ gauge rotations. The subgroup $\overline{\text{Aut}}$ of $\text{Aut}$ preserving the form (\ref{KS Ashtekar variables}) of the phase space variables $(A_a^i,E^a_i)$ yields a well-defined action on the parameters $(b,c,p_b,p_c)$ via 
\begin{align}
\label{phase space action}
\varphi \triangleright \left((A_a^i, E^a_i)(b,c,p_b,p_c)\right)
=: (A_a^i, E^a_i)(\varphi \triangleright(b,c,p_b,p_c))
\end{align}
for all $\varphi \in \overline{\text{Aut}}$. The quotient $\text{Aut}_R$ of $\overline{\text{Aut}}$ by the kernel of this action we call the group of \textit{residual automorphisms} in the KS framework. The identity component of $\text{Aut}_R$ consists of spatial diffeomorphisms, and we refer to it as the group of \textit{residual diffeomorphisms} for KS, and denote it $\text{Diff}_R$. 
If we let $\overline{\text{Diff}}$ denote the subgroup of $\overline{\text{Aut}}$ consisting in spatial diffeomorphisms, then 
$\text{Diff}_R$ can also be calculated as the quotient of $\overline{\text{Diff}}$ by the kernel of its action in \eqref{phase space action}.
In the present section we solve for the group $\text{Diff}_R$. The remaining discrete elements of $\text{Aut}_R$, which consist in the parity maps and their compositions, will be discussed in section \ref{Discrete Symmetries}. 

To solve for $\text{Diff}_R$, we first solve for $\overline{\text{Diff}}$ by finding the most general one parameter family of diffeomorphisms $s \mapsto \Phi_{\vec{v}}^s$, generated by some smooth vector field $\vec{v}$, preserving the form \eqref{KS Ashtekar variables}, so that 
\begin{align}
\Phi_{\vec{v}}^s \triangleright \left((A_a^i, E^a_i)(b,c,p_b,p_c)\right)
=: (A_a^i, E^a_i)(b(s),c(s),p_b(s),p_c(s))
\end{align}
for some set of functions $(b(s),c(s),p_b(s),p_c(s))$. Taking the derivative of both sides with respect to $s$ yields
    \begin{align}
    \label{reduced v flow}
    \begin{split}
            \mathcal{L}_{\vec{v}} A^i_a(s) &= \dot{A}^i_a = \frac{\partial A^i_a}{\partial b} \dot{b}(s) + \frac{\partial A^i_a}{\partial c} \dot{c}(s), %\nonumber 
            \\
            \mathcal{L}_{\vec{v}} E^a_i(s) &= \dot{E}^a_i = \frac{\partial E^a_i}{\partial p_b} \dot{p_b}(s) + \frac{\partial E^a_i}{\partial p_c} \dot{p_c}(s) . %\nonumber
            \end{split}
    \end{align}     
The set of all $\vec v$, satisfying these relations for some $\dot b$, $\dot c$, $\dot p_b$, and $\dot p_c$, will then generate $\overline{\text{Diff}}$. We proceed to derive the consequences of each of these conditions in the order most convenient: 
\begin{itemize}
    \item $A_a^2$:
\begin{align*}
    \LL_v A^2_a = v^b\del_b \left(b\del_a\theta\right) + b\del_b\theta\del_a v^b = b\left(\frac{\del v^\theta}{\del x}\del_ax + \frac{\del v^\theta}{\del \theta}\del_a\theta + \frac{\del v^\theta}{\del \phi}\del_a\phi\right),
\end{align*}
which must be equal to $\dot A_a^2 = \dot b\del_a\theta$, yielding 
\begin{align*}
    \frac{\del v^\theta}{\del x}=\frac{\del v^\theta}{\del\phi} = 0 \hspace{0.5cm} \textrm{and}\hspace{0.5cm}\dot b(s) = b\frac{\del v^\theta}{\del \theta}.
\end{align*}
Since $b$ and $\dot b$ are constant in space, so is $\frac{\del v^\theta}{\del \theta}$, which, together with the first two equations above, implies $v^\theta = \kappa_\theta\theta + \xi_\theta$ for some $\kappa_\theta, \xi_\theta \in \mathbb{R}$.   
However, smoothness of $\vec v$ requires that $v^\theta = 0$ at $\theta = 0$ and $\theta = \pi$, forcing $\kappa_\theta = \xi_\theta = 0$, whence
\begin{align}\label{cond vtheta}
    v^\theta \equiv 0.
\end{align}

\item $A_a^1$: 
        \begin{align*}
    \LL_{\vec{v}}A_a^1 & = v^b\del_b\left(-b\sin\theta\del_a\phi\right) - b\sin\theta\left(\del_b\phi\right)\del_av^b \\
    & = -bv^b\cos\theta\del_b\theta\del_a\phi -b\sin\theta\del_a v^\phi  \\
    & = -b\left(\cos\theta v^\theta +\sin\theta\frac{\del v ^\phi}{\del\phi}\right)\del_a\phi - b\sin\theta\frac{\del v^\phi}{\del\theta}\del_a\theta - b\sin\theta\frac{\del v^\phi}{\del x}\del_a x,
     \end{align*}
which, by \eqref{reduced v flow}, must be equal to $\dot A_a^1 = -\dot b(s)\sin\theta\partial_a\phi$. This, with \eqref{cond vtheta}, implies
\begin{align*}
  \frac{\del v^\phi}{\del\theta} = \frac{\del v^\phi}{\del x} = 0 \hspace{0.5cm} \textrm{and}\hspace{0.5cm} \dot b(s) = b\frac{\del v^\phi}{\del\phi}.  
\end{align*}
Thus, by the same argument used for $v^\theta$, we conclude $v^\phi = \kappa_\phi\phi + \xi_\phi$ for some constants $\kappa_\phi, \xi_\phi \in \mathbb{R}$. Smoothness of $\vec v$ now requires $v^\phi(\phi = 0) = v^\phi(\phi = 2\pi)$, forcing $\kappa_\phi = 0$, so that
\begin{align}\label{cond vphi}
    v^\phi = \text{constant} &=: \xi_\phi .
\end{align}

\item $A_a^3$:
            \begin{align*}
            \LL_{\vec{v}}A_a^3 &= v^b\del_b\left( \frac{c}{L_0}\del_a x + \cos\theta\del_a\phi\right) + \left(\frac{c}{L_0}\del_b x + \cos\theta\del_b\phi\right)\del_a v^b \nonumber \\
            & = -\sin\theta v^\theta\del_a\phi +  \frac{c}{L_0}\del_av^x + \cos\theta\phi\del_a v^\phi \\
            & =  \frac{c}{L_0}\frac{\del v^x}{\del\phi}\del_a\phi  + \frac{c}{L_0}\frac{\del v^x}{\del \theta}\del_a\theta  + \frac{c}{L_0}\frac{\del v^x}{\del x} \del_ax     
            \end{align*}
where, in going from the second to the third line, we used \eqref{cond vtheta} and \eqref{cond vphi}. Requiring this to be equal to $\dot A_a^3 = \frac{\dot c}{L_0}\del_ax$ then implies  
\begin{align}
    \label{flow A3}
    \frac{\del v^x}{\del \phi} = \frac{\del v^x}{\del\theta}= 0 \hspace{0.5cm} \hspace{0.5cm}\textrm{and}\hspace{0.5cm} \dot c = c \frac{\del v^x}{\del x}.
\end{align}
Since $c$ and $\dot c$ are constant in space, the same argument used for $v^\theta$ and $v^\phi$ again applies here, so that 
\begin{align}\label{cond vx}
    v^x = \kappa_x x + \xi_x 
\end{align}
for some constants $\kappa_x, \xi_x \in \mathbb{R}$, this time unconstrained by smoothness of $\vec{v}$. 
\end{itemize}
The restrictions \eqref{cond vtheta} to \eqref{cond vx} thereby obtained fix 
\begin{equation}\label{v}
    \vec{v} = \xi_\phi\vec\phi + (\xi_x + \kappa_xx)\vec{x},
\end{equation}
where $\xi_\phi, \xi_x, \kappa_x$ are free constant parameters.
One can check that the remaining conditions in \eqref{reduced v flow} are automatically satisfied with no further restriction on $\vec{v}$ --- explicitly, from 
$\LL_v E^a_i = v^c \partial_c E^a_i - E^c_i \partial_c v^a +E^a_i \partial_c v^c$, 
\begin{align}
    \label{flow E1} \LL_v E^a_1 & =-\frac{\kappa_xp_b}{L_0}\phi^a, \\
    \label{flow E2} \LL_v E^a_2 & = \frac{\kappa_x p_b}{L_0}\sin\theta \theta^a, \\
    \label{flow E3} \LL_v E^a_3 &= 0 .
\end{align}
Therefore $\left\{\vec\phi,\vec x, x \vec x\right\}$ is a basis of the vector fields generating $\overline{\text{Diff}}$.
Note that the resulting flows --- \eqref{flow A3}, \eqref{cond vx}, \eqref{flow E1} and \eqref{flow E2} --- depend only on $\kappa_x$, and not on $\xi_x$ or $\xi_\phi$. The reason why is easy from the significance of the corresponding vector fields: 
\begin{itemize}
    \item $\vec{\phi}$ generates part of the spherical symmetry manifest in Schwarzschild. The other two spatial rotations are not manifest here as symmetries because we are looking at symmetries of $(A_a^i,E^a_i)$ --- full spherical symmetry can be imposed on $(A_a^i,E^a_i)$ at most up to $SU(2)$ gauge rotations, and is manifest only in $SU(2)$-gauge invariant structures constructed from them, such as the 3-metric (\ref{KS metric}).
    
    \item $\vec{x}$ generates translations in $x$, which corresponds to $t$ in the usual form of the Schwarzschild solution, so that this symmetry corresponds to the $t$-translation symmetry in Schwarzschild. 
    
    \item $x\vec{x}$ generates something more interesting: An exponential flow in the $x$ direction, and the only flow with non-trivial action on $(A_a^i, E^a_i)$. 
\end{itemize}
Thus, the kernel $K$ of the action of $\overline{\text{Diff}}$ on $(A_a^i,E^a_i)$ is generated by $\vec{\phi}$ and $\vec{x}$, 
so that the group of residual diffeomorphisms $\text{Diff}_R:=\overline{\text{Diff}}/K$ is one dimensional, 
parameterized by $\kappa_x$. 
Rescaling $\vec{v}$ in \eqref{v} is equivalent to rescaling the parameter time $s$ for the flow generated, so that we can, without loss of generality, take $\kappa_x = 1$. With this choice, the resulting flow of the phase space variables $(b,p_b,c,p_c)$ is given by
\begin{align}\label{resulting flows}
    \dot b  = 0, \qquad
    \dot p_b = p_b, \qquad
    \dot c = c, \qquad
    \dot p_c = 0.
\end{align}
The volume of the fiducial cell then flows as $\dot V = 4\pi\dot{|p_b|}\sqrt{|p_c|} = 4\pi|p_b|\sqrt{|p_c|} = V$, and hence the flow of the Hamiltonian constraint is of the form 
    \begin{align}
        \label{Hcl flow} \dot H_{c\ell} = (n+1)H_{c\ell}.
    \end{align}

\section{Covariance Equation}\label{Covariance Equation}

\subsection{Strategy}

Classically, the flow of a phase space function $F$ under a family of canonical transformations generated by phase space function $\Lambda$ is given by
\begin{align}\label{canonical transf.}
    \dot{F} = \{\Lambda,F\}.
\end{align}
Standard quantization procedure then turns functions into operators and Poisson brackets into commutators, yielding the following evolution with respect to the flow parameter $s$:
\begin{align*}
    \dot{F} = \frac{1}{i\hbar}\left[\hat{\Lambda},\hat{F}\right] \hspace{0.5cm} \Rightarrow \hspace{0.5cm}    \hat F(t) = e^{\frac{s}{i\hbar}\hat\Lambda}\hat F(0) e^{-\frac{s}{i\hbar}\hat\Lambda}.
\end{align*}
The residual diffeomorphism flow in Kantowski-Sachs, however, does not preserve Poisson brackets and so is non-canonical. The flow can, however, be cast in a form related to (\ref{canonical transf.}),
\begin{align}\label{Non-canonical transf}
    \dot F & = \omega_1(b,p_b)\{\Lambda_1(b,p_b),F\} + \omega_2(c,p_c)\{\Lambda_2(c,p_c),F\}.
\end{align}
Equations \eqref{resulting flows} then become
\begin{align*}
    0 = -\gamma G \omega_1\frac{\del \Lambda_1}{\del p_b}, \quad
    p_b = \gamma G \omega_1 \frac{\del \Lambda_1}{\del b}, \quad
    c = -2\gamma G \omega_2\frac{\del \Lambda_2}{\del p_c}, \quad
    0 = 2\gamma G \omega_2\frac{\del \Lambda_2}{\del c}.
\end{align*}
The first and last equations tell us that $\Lambda_1 = \Lambda_1(b)$ and $\Lambda_2 = \Lambda_2(p_c)$ are each a function of only one variable. The remaining equations then determine $\omega_{1}$ and $\omega_2$ in terms of $\Lambda_1$ and $\Lambda_2$,
\begin{align*}
    \omega_1(b,p_b) = \frac{p_b}{\gamma G \frac{\del \Lambda_1(b)}{\del b}}, \hspace{0.5cm} \omega_2(c,p_c) = -\frac{c}{2\gamma G\frac{\del \Lambda_2(p_c)}{\del p_c}}.
\end{align*}
Therefore, the only free parameters are $\Lambda_1(b)$ and $\Lambda_2(p_c)$, with a restriction that their first derivatives do not vanish, except possibly on a set of measure zero. Arguably, the simplest choice is to make $\Lambda_1(b)$ and $\Lambda_2(p_c)$ proportional to $b$ and $p_c$, respectively. The choice of proportionality constant does not effect the final quantum covariance condition, and so, without loss of generality, we set
\begin{align}\label{solution 1}
    \Lambda_1 =  b & \hspace{0.5cm} \Rightarrow \hspace{0.5cm} \omega_1 = \frac{p_b}{\gamma G }. \nonumber \\ \Lambda_2 = p_c & \hspace{0.5cm} \Rightarrow \hspace{0.5cm} \omega_2 = - \frac{c}{2\gamma G}.
\end{align}
With this choice, equation \eqref{Hcl flow} takes the form 
\begin{align}
\label{pbflow}
    \dot H_{c\ell} = \frac{p_b}{\gamma G}\left\{b, H_{c\ell}\right\} - \frac{c}{2\gamma G}\left\{p_c,H_{c\ell}\right\} = (n+1) H_{c\ell} .
\end{align}

It is equation \eqref{pbflow} that we will quantize to obtain a quantum covariance condition on the constraint operator $\hat H$. Since $b$ and $c$ appear directly in this equation without exponentiation, and since $\hat b$ and $\hat c$ are not well defined on the Bohr Hilbert space arising from loop quantization (as reviewed in section \ref{Quantum Kinematics}), we first find the general solution to this equation in the standard Schr\"odinger representation, with subsequent imposition of preservation of the Bohr Hilbert space.

\subsection{Quantization in the Schr\"odinger representation and general solution for the matrix elements}
\label{quantum covariance section}

We follow the standard quantization procedure, choosing the Weyl ordering for quantizing products, $\hat A \star \hat B := \frac{1}{2}\left(\hat A\hat B + \hat B\hat A\right)$, yielding
\begin{align}\label{HWeyl}
    (n+1)\hat H &= \frac{1}{i\hbar}\left( \hat{\omega}_1\star\left[\hat\Lambda_1,\hat H\right] + \hat{\omega}_2\star\left[\hat{\Lambda}_2,\hat H\right]\right) \nonumber \\
    & = \frac{1}{2i\gamma\ell_P^2 }\left(\hat p_b\left[\hat b,\hat H\right] + \left[\hat b,\hat H\right]\hat p_b\right) - \frac{1}{4i\gamma \ell_P^2}\left(\hat c\left[\hat p_c,\hat H\right] + \left[\hat p_c,\hat H\right]\hat c\right).
    \end{align}
The strategy is to recast (\ref{HWeyl}) in terms of the matrix elements of $\hat H$ in the $|p_b',p_c'\rangle$ basis, where the action of the position operators is given by
% \begin{align}
%     \hat b = i\gamma \ell_P^2 \frac{\del}{\del p_b}, \hspace{0.5cm} \hat c = 2i\gamma \ell_P^2 \frac{\del}{\del p_c},
% \end{align}
% which are equivalent to the identities
\begin{align*}
    \langle p_b',p_c'|\hat b = i\gamma\ell_P^2\frac{\del}{\del p_b'}\langle p_b',p_c'| \nonumber \\
    \langle p_b',p_c'|\hat c = 2i\gamma\ell_P^2\frac{\del}{\del p_c'}\langle p_b',p_c'|,
\end{align*}
and their conjugates. We have
\begin{align*}
(n+1)\langle p_b'',p_c''|\hat{H}|p_b',p_c'\rangle & = 
\frac{1}{2i\gamma\ell_P^2}\left\langle p_b'',p_c''\left|\hat{p}_b\hat b\hat H - \hat{p}_b\hat H\hat b +\hat b\hat H\hat p_b -\hat H\hat b\hat p_b \right|p_b',p_c'\right\rangle \nonumber \\
&\ \ \ - \frac{1}{4i\gamma\ell_P^2}\left\langle p_b'',p_c''\left|\hat c\hat p_c\hat H-\hat c\hat H\hat p_c +\hat p_c\hat H\hat c -\hat H\hat p_c\hat c \right|p_b',p_c'\right\rangle \nonumber 
\end{align*}
\begin{align}
\label{matrix equation}
\rule{0in}{0in} \hspace{1em}& = \frac{1}{2}\left(p_b''\frac{\del}{\del p_b''} - p_b''\left(-\frac{\del}{\del p_b'}\right)+\frac{\del}{\del p_b''}p_b'-p_b'\left(-\frac{\del}{\del p_b'}\right)\right) \left\langle p_b'',p_c''|\hat H|p_b',p_c'\right\rangle \nonumber \\
&\ \  - \frac{1}{2}\left(\frac{\del}{\del p_c''}p_c'' - \frac{\del}{\del p_c''}p_c' + p_c'' \left(-\frac{\del}{\del p_c'}\right) - \left(-\frac{\del}{\del p_c'}\right)p_c'\right) \left\langle p_b'',p_c''|\hat H|p_b',p_c'\right\rangle \nonumber \\
% & = \frac{1}{2}\left\{p_b''\frac{\del}{\del p_b''} + p_b''\frac{\del}{\del p_b'}+\frac{\del}{\del p_b''}p_b'+p_b'\frac{\del}{\del p_b'}\right\} \left\langle p_b'',p_c''|\hat H|p_b',p_c'\right\rangle \nonumber \\
% &\ \ - \frac{1}{2}\left\{1 + p_c''\frac{\del}{\del p_c''} - p_c'\frac{\del}{\del p_c''} - p_c'' \frac{\del}{\del p_c'} + 1 + p_c'\frac{\del}{\del p_c'}\right\} \left\langle p_b'',p_c''|\hat H|p_b',p_c'\right\rangle \nonumber \\
& = \left(\frac{1}{2}\left(p_b' + p_b''\right)\left(\frac{\del}{\del p_b'} + \frac{\del}{\del p_b''}\right) - \frac{1}{2}\left(p_c'' - p_c'\right)\left(\frac{\del}{\del p_c''} - \frac{\del}{\del p_c'}\right) - 1 \right)\cdot \nonumber \\
& \rule{0in}{0in} \hspace{3in} \cdot \left\langle p_b'',p_c''|\hat H|p_b',p_c'\right\rangle. 
\end{align}
% Thus
% \begin{align}
%     \left\{\frac{1}{4}\left(p_b' + p_b''\right)\left(\frac{\del}{\del p_b'} + \frac{\del}{\del p_b''}\right) - \frac{1}{4}\left(p_c'' - p_c'\right)\left(\frac{\del}{\del p_c''} - \frac{\del}{\del p_c'}\right) \right\}\left\langle p_b'',p_c''|\hat H|p_b',p_c'\right\rangle = \left\langle p_b'',p_c''|\hat H|p_b',p_c'\right\rangle.
% \end{align}
Making the change of variables
\begin{align*}
    u_b = p_b' + p_b'', &\hspace{0.5cm}  v_b = p_b'' - p_b' \nonumber \\
    u_c = p_c'+p_c'', &\hspace{0.5cm} v_c = p_c'' - p_c'
\end{align*}
and defining
\begin{align*}
f(u_b,v_b,u_c,v_c):= \left\langle p_b'',p_c''|\hat H|p_b',p_c'\right\rangle,
\end{align*}
this becomes
\begin{align}\label{matrix H equation}
   \left(u_b\frac{\del}{\del u_b} - v_c\frac{\del}{\del v_c}\right) f(u_b,v_b,u_c,v_c) = (n+2) f(u_b,v_b,u_c,v_c) .
\end{align}
Now, for a general path $(u_b(s),v_b(s),u_c(s),v_c(s))$ in the parameter space, we have
\begin{align*}
    \frac{df}{ds} = \frac{\del f}{\del u_b}\frac{d u_b}{d s} + \frac{\del f}{\del v_b}\frac{d v_b}{d s} + \frac{\del f}{\del u_c}\frac{d u_c}{d s} + \frac{\del f}{\del v_c}\frac{d v_c}{d s} .
\end{align*}
As long as the chosen path satisfies $\frac{d u_b}{ds} = u_b$,
$\frac{du_c}{ds}, \frac{dv_b}{ds} = 0$, and $\frac{dv_c}{ds} = -v_c$,
which is equivalent to 
\begin{align*}
    &u_b(s) = C_be^s,
    \text{ for some }C_b \in \mathbb{R},
     \\
    &v_c(s) = C_ce^{-s},
    \text{ for some }C_c \in \mathbb{R},\text{ and}
     \\
    &u_c, v_b \in \mathbb{R},\text{ constant},
\end{align*}
then equation \eqref{matrix H equation} then simplifies
to
\begin{align*}
    \frac{d f}{ds} &= (n+2)f 
\end{align*}
with general solution
\begin{align*}
    f &= C e^{(n+2)s} .
\end{align*}
That is, for all $C_b, v_b, u_c, C_c \in \mathbb{R}$, there
exists some real $C(C_b, v_b, u_c, C_c)\in \mathbb{R}$ such that 
\begin{align}
\label{general s}
    f(C_b e^s, v_b, u_c, C_c e^{-s}) = C(C_b, v_b, u_c, C_c) e^{(n+2)s}
\end{align}
for all $s \in \mathbb{R}$. In particular, for $s = -\ln |C_b|$, this becomes
\begin{align*}
    f(\sgn(C_b), v_b, u_c, C_c |C_b| ) = C(C_b, v_b, u_c, C_c) |C_b|^{-(n+2)}.
\end{align*}
Using this to eliminate $C(C_b, v_b, u_c, C_c)$ from \eqref{general s}, and letting $u_b := C_b e^s$ and 
$v_c := C_c e^{-s}$, we obtain
\begin{align}
\label{f solution}
    f(u_b, v_b, u_c, v_c) = 
    C_{\sgn(u_b)}(v_b, u_c, v_c |u_b|) |u_b|^{n+2} ,
\end{align}
where we have defined
\begin{align*}
   C_{\sigma}(v_b, u_c, w):= f(\sigma, v_b, u_c, w).
\end{align*}
Using the fact that we have assumed $n>-3$, one can check that \eqref{f solution} satisfies \eqref{matrix H equation} with no further restriction. 
%
% The n > -3 restriction is required for the term
% \delta(u_b)|u_b|^{n+3} appearing on the left hand side of \eqref{matrix H equation} to vanish.
%
That is, from \eqref{f solution}, the general solution to equation \eqref{matrix equation} is 
\begin{align}
    \label{matrix H solution}
    \left\langle p_b'',p_c''|\hat H|p_b',p_c'\right\rangle & = 
    C_{\sgn(p_b''+p_b')}\left(p_b'' - p_b', p_c'' + p_c',|p_b''+p_b'|(p_c'' - p_c')\right)|p_b''+p_b'|^{n+2}
\end{align}
with $C_\sigma(v_b, u_c, w)$ arbitrary and real.

\subsubsection{Operator form of the solution}

From (\ref{matrix H solution}) the action of $\hat H$ on an arbitrary state $|p_b',p_c'\rangle$ is
\begin{align}\label{H integral1}
    \hat H |p_b',p_c'\rangle &= \int |p_b'',p_c''\rangle\langle p_b'',p_c''|\hat H|p_b',p_c'\rangle dp_b''dp_c'' \nonumber \\
    & = \int C_{\sgn(p_b''+p_b')}\left(p_b'' - p_b', p_c'' + p_c',|p_b''+p_b'|(p_c'' - p_c')\right)|p_b''+p_b'|^{n+2} |p_b''p_c''\rangle dp_b''dp_c''.
\end{align}
Define new variables $A$ and $B$ by
\begin{align}\label{shifts}
            p_b'' &= p_b' + (p_b'' - p_b') =: p_b' + \gamma\ell_P^2A \nonumber \\ p_c'' &= p_c' + \frac{\frac{1}{2}|p_b'+p_b''|(p_c''-p_c')}{|p_b' +\frac{1}{2}(p_b'' - p_b')|} =: p_c' + \frac{4\gamma\ell_P^4B}{|p_b'+\frac{1}{2}\gamma\ell_P^2A|}
        \end{align}
where the reason for this definition will be clear in further steps. Performing the change of variables from $(p_b'',p_c'')$ to $(A,B)$ in the integral \eqref{H integral1} gives
% we have
% \begin{align*}
%     \label{Jacobian AB} \left|\det\left(\begin{matrix} \frac{\del p_b''}{\del A} & \frac{\del p_c''}{\del A} \\ \frac{\del p_b''}{\del B} & \frac{\del p_c''}{\del B} \end{matrix}\right)\right| = \left|\det\left(\begin{matrix} \gamma \ell_P^2 & \frac{2\gamma^2\ell_P^6B}{(p_b'+\frac{1}{2}\gamma\ell_P^2A)^2} \\ 0 & \frac{4\gamma\ell_P^4}{|p_b'+\frac{1}{2}\gamma\ell_P^2A|}\end{matrix}\right)\right| = \frac{4\gamma^2\ell_P^6}{|p_b'+\frac{1}{2}\gamma\ell_P^2A|}
% \end{align*} and 
%(\ref{H integral1}) takes the form
\begin{align}\nonumber
     \hat H |p_b',p_c'\rangle =
     %\int C_{\sgn(2p_b' + \gamma\ell_P^2A)}\left[A,2p_c' + \frac{4\gamma\ell_P^4B}{p_b'-\frac{1}{2}\gamma\ell_P^2A}, B\right]8\gamma^2\ell_P^6\sgn(2p_b' + \gamma\ell_P^2A)\left(2p_b' + \gamma\ell_P^2A\right)^{n+1}\left|p_b' + \gamma\ell_P^2A, p_c' + \frac{4\gamma\ell_P^4B}{p_b'-\frac{1}{2}\gamma\ell_P^2A}\right\rangle dAdB = \nonumber \\
    &\int C'_{\sgn(2p_b' + \gamma\ell_P^2A)}\left(A,2p_c' + \frac{4\gamma\ell_P^4B}{|p_b'+\frac{1}{2}\gamma\ell_P^2A|}, B\right)|2p_b' + \gamma\ell_P^2A|^{n+1} \cdot \\
    &\cdot \left|p_b' + \gamma\ell_P^2A, p_c' + \frac{4\gamma\ell_P^4B}{|p_b'+\frac{1}{2}\gamma\ell_P^2A|}\right\rangle dAdB 
    \label{H integral A,B}
    %\nonumber \\     &= \left[\int {e^{\frac{iA}{2}\hat b}}{e^{\frac{iB}{2}\frac{\hat c}{\hat p_b}}}\ \hat p_b\ \alpha(A,B,\hat p_c,\sgn p_b){e^{\frac{iB}{2}\frac{\hat c}{\hat p_b}}}{e^{\frac{iA}{2}\hat b}}dAdB\right]|p_b',p_c'\rangle
\end{align}
where
$C_\sigma'(A,u_c,B) := 8\gamma^2\ell_P^2 C_\sigma(\gamma \ell_P^2 A, u_c, 8\gamma\ell_P^4 B)$. This result can then be written by using the action of shift operators, as 
\begin{align} \nonumber
   &\hat H |p_b',p_c'\rangle  = \\
   \label{action H integral}
   &\left(\int {e^{\frac{iA}{2}\hat b}}{e^{\frac{iB}{2}\frac{\hat c}{|\hat p_b|}}}\ |\hat p_b|^{n+1} \alpha(A,B,\hat p_c,\sgn p_b){e^{\frac{iB}{2}\frac{\hat c}{|\hat p_b|}}}{e^{\frac{iA}{2}\hat b}}dAdB\right)|p_b',p_c'\rangle
\end{align}
for $\alpha:\mathbb{R}^3\times\{\pm 1\}\rightarrow \mathbb{C}$ an unconstrained parameter function related to $C'$.
If we define the following quantization prescription for any function 
$f(p_b,p_c)$,
\begin{align}
    \label{operator Frankstein} \reallywidehat{ f(p_b,p_c)e^{i\left(Ab + B\frac{c}{|p_b|}\right)}} : =   e^{\frac{iA}{2}\hat b}e^{\frac{iB}{2}\frac{\hat c}{|\hat p_b|}}  f(\hat p_b, \hat p_c)e^{\frac{iB}{2}\frac{\hat c}{|\hat p_b|}}e^{\frac{iA}{2}\hat b},
\end{align}
then the Hamiltonian constraint operator takes the form
\begin{align}\label{H constraint integral}
    \hat H = \int\ \reallywidehat{|p_b|^{n+1} \alpha(A,B, p_c,\sgn p_b)e^{i\left(Ab + B\frac{c}{|p_b|}\right)}}\ dAdB.
\end{align}

\subsection{Preservation of Bohr Hilbert space} 

Until now we have been working in the Schr\"odinger representation of the quantum algebra of kinematical observables. However, the representation descending from full loop quantum gravity, and in the simpler isotropic case selected by residual diffeomorphism covariance \cite{Engle:2016hei,Engle:2016zac}, is the representation on the Bohr Hilbert space. 
In order to ensure that the operator $\hat H$ is well-defined on this Hilbert space, it must preserve at least a subset of it dense with respect to its inner product. More precisely, we require that $\hat H$ map at least one finite linear combination of momentum eigenstates back into the Bohr Hilbert space, so that, in particular, for any $p_b'',p_c''$ there is at most countable $p_b',p_c'$ such that the matrix elements (\ref{matrix H solution}) are non-zero. 
This will be true if and only if the function $\alpha$ appearing in \eqref{H constraint integral} is an at most countable sum of Dirac delta functions over the integration variables $A, B$,
    \begin{align}\label{integral into sum}
        \alpha(A,B,p_c,\sgn p_b) = \sum_k \alpha_k(p_c, \sgn p_b)\delta(A - A_k(p_c))\delta(B - B_k(p_c))
    \end{align}
where the peaks of the Dirac delta functions are allowed to depend on the third continuous parameter, $p_c$. 
The Hamiltonian operator then takes the form
    \begin{align}
        \label{H preserved Bohr} 
        \hat H = \sum_k \reallywidehat{|p_b|^{n+1} \alpha_k(p_c,\sgn p_b)e^{i\left(A_k(p_c)b + B_k(p_c)\frac{c}{|p_b|}\right)}}.
    \end{align}

\section{Discrete Symmetries} \label{Discrete Symmetries}

    The form \eqref{H preserved Bohr} for the quantum Hamiltonian is the most general that is covariant under the one parameter family of residual diffeomorphisms. The remaining discrete residual automorphisms of the $SU(2)$ principal bundle are parity maps which preserve the Poisson brackets of the classical theory and so correspond to unitary transformations in the quantum theory. Explicitly:
    \begin{description}
    \item \textit{`b-parity'} $\Pi_b:(b,p_b)\mapsto(-b,-p_b)$ is equivalent to an internal gauge rotation of $\pi$ around the 3-axis, with corresponding quantum map given by
    $\hat{\Pi}_b|p_b',p_c'\rangle := |-p_b',p_c'\rangle$. 
    \item \textit{`c-parity'}
    $\Pi_c: (c,p_c)\mapsto (-c,-p_c)$ is equivalent to the action of the antipodal map $(\theta,\phi)\mapsto(\pi-\theta,\phi+\pi)$ as a diffeomorphism, combined with internal parity along the 3-axis, with corresponding quantum map given by $\hat{\Pi}_b|p_b',p_c'\rangle := |-p_b',p_c'\rangle$. 
    \end{description}
    The classical Hamiltonian $H_{c\ell}$ is odd under both of these parities, so that we likewise impose that 
    the quantum Hamiltonian $\hat H$ be odd under conjugation by the corresponding unitary operators. This, together with condition that $\hat H$ be invariant under Hermitian conjugation, make up the discrete symmetries to impose on $\hat H$.
    
    We define the \textit{classical analogue} of the operator $\hat H$ in \eqref{H preserved Bohr} to be its preimage under our quantization map, namely
    \begin{align}\label{classical analogue}
         H = \sum_k |p_b|^{n+1} \alpha_k(p_c, \sgn p_b)e^{i\left(A_k(p_c)b + B_k(p_c)\frac{c}{|p_b|}\right)}.
    \end{align}
    It is somewhat remarkable and convenient that our quantization map \eqref{operator Frankstein}, which was naturally suggested by the solution to the quantum residual diffeomorphism covariance condition, additionally (1.) intertwines complex conjugation and Hermitian conjugation ($\hat H^\dagger = \hat{\bar H}$) and (2.) is covariant with respect to the parity maps
    ($\hat\Pi_b\hat H\hat\Pi_b = \reallywidehat{\Pi_b^*H}$, $\hat\Pi_c\hat H\hat\Pi_c = \reallywidehat{\Pi_c^*H}$). As a consequence, imposing that $\hat H$ be Hermitian and covariant under the quantum parity maps is equivalent to imposing that the classical analogue $H$ \eqref{classical analogue} be real and covariant under the classical parity maps. The most general such $H$ 
    can always be cast in the form
    \begin{align}
        \nonumber
        H =& |p_b|^{n+1}\sgn (p_bp_c)a_0(p_c) \; + \; |p_b|^{n+1}\sum_{k>0}\bigg(\alpha_k(p_c,\sgn p_b)e^{i\left(A_k(p_c)b + B_k(p_c)\frac{c}{|p_b|}\right)} \\
        & \rule{0in}{0in}\hspace{4em} - \,\,\left((b,p_b)\!\!\mapsto\!\!(-b,-p_b)\right)\,\,
        - \,\,\left((c,p_c)\!\!\mapsto\!\!(-c,-p_c)\right)\,\,
        +\,\, c.c \bigg)  
        \label{H most general}
    \end{align}
    where $a_0(p_c)$ is even and real, $(b,p_b)\mapsto(-b,-p_b)$ denotes the foregoing terms in the large parenthesis with the indicated replacement, $(c,p_c)\mapsto(-c,-p_c)$ similarly, and $c.c.$ denotes the complex conjugate of the foregoing terms, so that the number of terms in the large parentheses is eight.  Note that, compared to the form \eqref{classical analogue}, the terms above have been relabeled so that each label $k>0$ corresponds to eight terms, for convenience.

\paragraph{Metric loop assumption:}

As discussed in section \ref{Quantum Kinematics}, the functions of the connection with direct quantum analogues are parallel transports along paths. The Hamiltonian constraint is linear in the curvature of the connection, which must therefore be quantized by first regularizing the curvature in terms of holonomies around loops. In minisuperspace quantizations such as the present one, the limit in which these loops approach a point is taken by choosing the loops so that they enclose an area equal to the minimal non-zero eigenvalue $\Delta$ of the area operator in full loop quantum gravity. As a consequence, the choice of loops depends on the triad; however, more specifically, it depends on the \textit{metric} determined by the triad. Thus, in the resulting expression for the holonomies, and hence the regularized constraint, the coefficients $A_k(p_c)$ and $B_k(p_c)$ of the connection components must be even. We call this assumption the \textit{metric loop assumption}.
The consequent symmetry of the coefficients $A_k(p_c)$ and $B_k(p_c)$ is the final discrete symmetry we consider.

With this assumption, it becomes convenient to decompose each coefficient $\alpha_k(p_c, \sgn p_b)$ into its even and odd parts in each argument, as well as into its real and imaginary parts,
\begin{align}
\nonumber
\alpha_k(p_c,\sgn p_b) =: & \frac{1}{8}\bigg(\Big(a_k(p_c) + i\tilde{a}_k(p_c)\Big)\sgn(p_b p_c)
-\Big(\tilde{b}_k(p_c)+ib_k(p_c)\Big)\sgn p_b \\
& - \Big(\tilde{c}_k(p_c)+ic_k(p_c)\Big)\sgn p_c - d_k(p_c)-i\tilde{d}_k(p_c)
\bigg),
\end{align}
with $a_k, \tilde{a}_k, b_k, \tilde{b}_k, c_k, \tilde{c}_k, d_k, \tilde{d}_k$ real and even functions of $p_c$. The terms in the summand in \eqref{H most general} then reduce to only four terms involving sines and cosines, yielding the following more explicit form: 
\begin{align} \label{H final}
        \nonumber
        H =&  |p_b|^{n+1} a_0 \sgn (p_bp_c) \; + \; |p_b|^{n+1}\sum_{k>0}\bigg(
        a_k \sgn(p_b p_c) \cos(A_k b)\cos\left(B_k \frac{c}{|p_b|}\right) \\
        \nonumber
        &\rule{0in}{0in} \hspace{3em} + b_k \sgn(p_b) \cos(A_k b)\sin\left(B_k \frac{c}{|p_b|}\right) 
        + c_k \sgn(p_c) \sin(A_k b)\cos\left(B_k \frac{c}{|p_b|}\right) \\
        &\rule{0in}{0in} \hspace{3em} + d_k \sin(A_k b)\sin\left(B_k \frac{c}{|p_b|}\right)\bigg)   
    \end{align}
with $a_k, b_k, c_k, d_k, A_k, B_k$ (thus far arbitrary) even functions of $p_c$ alone.

\section{Classical asymptotic behaviour}\label{Classical asymptotic behavior}

\subsection{Na\"ive classical limit and the limit of low curvature}
\label{low curvature}

% A necessary check of consistency for any model of quantum gravity is that it should match GR in regimes of low curvature, where the latter is well tested. Therefore, \eqref{H final} should match \eqref{Hcl} in such a limit.

The standard way to define the classical limit (and, indeed, the only used in the LQC and loop quantum KS literature so far) is to take the limit as the arguments of the exponentials (or sines) go to zero, which is related to an $\ell_P \rightarrow 0$ limit of such arguments \cite{Ashtekar:2018cay,Campiglia:2007pb, Chiou:2008eg,Joe:2014tca,Cortez:2017alh,Bodendorfer:2019nvy, Assanioussi:2020ezr}. This limit is in fact equivalent to the limit in which the eigenvalues of extrinsic curvature go to zero. To see this, from equations \eqref{KS Ashtekar variables}, one can calculate
\begin{align*}
    % K_{ab} &= \frac{b\sin^2\theta\sqrt{|p_c|}}{\gamma\sgn{(p_b)}}\del_a\phi\del_b\phi + \frac{b\sqrt{|p_c|}}{\gamma\sgn{(p_b)}}\del_a\theta\del_b\theta
    % + \frac{c|p_b|}{\gamma L_0^2\sgn{(p_c)}\sqrt{|p_c|}}\del_ax\del_bx.
     % K &= K_{ab}q^{ab} = \frac{2b}{\gamma\sgn(p_b)\sqrt{|p_c|}} + \frac{c\sqrt{|p_c|}}{\gamma\sgn(p_c)|p_b|}, 
     K_a^{\ b} &=  \frac{b}{\gamma\sgn(p_b)\sqrt{|p_c|}}\del_a\phi\phi^b + \frac{b}{\gamma\sgn(p_b)\sqrt{|p_c|}}\del_a\theta\theta^b + \frac{c\sqrt{|p_c|}}{\gamma\sgn(p_c)|p_b|}\del_axx^b,
\end{align*}
from which one can read off the eigenvalues of the extrinsic curvature, the limit of whose vanishing is then equivalent to the two conditions
\begin{align}\label{classicality}
        \mathfrak{b} := \frac{b}{\sqrt{|p_c|}} << \frac{1}{\ell_P}, \hspace{0.5cm}\textrm{and}\hspace{0.5cm} \mathfrak{c}:= \frac{\sqrt{|p_c|}c}{|p_b|} << \frac{1}{\ell_P},
\end{align}
which, for fixed $p_c$, are equivalent to the vanishing of the arguments of the exponentials in \eqref{H final}. 
As this is equivalent to the definition of the classical limit in all prior LQC and loop quantum KS literature,
and as the focus of this paper is on the consequences of residual diffeomorphism covariance, and not the development of a new condition for imposing the classical limit, this is the definition that we use here as well.  

However, we would like to point out that this condition is \textit{not sufficient}, because the true regime in which the correct classical limit should be imposed is that of small \textit{four dimensional curvature scalars}, a condition which can be satisfied without the extrinsic curvature being small. Indeed, this is what happens at the horizon in Kantowski-Sachs: The eigenvalues of the extrinsic curvature diverge, while four dimensional curvature scalars remain small compared to the Planck scale. In fact, we believe this is the reason why, up to now, $\overline{\mu}$-schemes have failed to have the correct classical limit at the horizon, something which the models \cite{Ashtekar:2018cay,Ashtekar:2023cod,Corichi:2015xia,Olmedo:2017lvt} improve upon, and which we remark upon further in section \ref{aossect}. 

Adapting to the limit \eqref{classicality},
one can rewrite the effective Hamiltonian \eqref{H final} in terms of $\mathfrak{b},\mathfrak{c}$
%and new coefficients $\mathcal{A}_k,\mathcal{B}_k$, 
by replacing 
\begin{align*}
    A_kb \mapsto \sqrt{|p_c|} A_k\bb  \hspace{1cm}\textrm{and}\hspace{1cm} 
    B_k\frac{c}{|p_b|} \mapsto \frac{1}{\sqrt{|p_c|}} B_k\cc.
\end{align*}
The classical limit is then obtained by considering the leading terms in the asymptotic expansion in the limit $\left(\bb,\cc\right)\to(0,0)$.

\subsection{Equations for correct asymptotic behavior in the na\"ive classical limit}

When comparing the expanded Hamiltonian with \eqref{Hcl}, one should ask which terms are relevant to contribute to $H_{c\ell}$, and which are subdominant. The classical Hamiltonian has a form $H_{c\ell} = A \cdot 1 +B\mathfrak{b}^2+C\mathfrak{b}\mathfrak{c}$, and relevance or subdominance relative each component must be checked separately. Specifically, for given $n,m$, if
\begin{align*}
\lim_{(\bb,\cc)\rightarrow (0,0)} \frac{\bb^n \cc^m}{1} \ =  \ \lim_{(\bb,\cc)\rightarrow (0,0)} \frac{\bb^n \cc^m}{\bb^2}\ =  \  \lim_{(\bb,\cc)\rightarrow (0,0)}  \frac{\bb^n \cc^m}{\bb\cc}\ =\ 0,
\end{align*}
independently of how the limit is taken, then $\bb^n \cc^m$ is subdominant to each term in $H_{cl}$ in the classical limit, otherwise we call the term \textit{relevant} and require the coefficients to match the corresponding ones in $H_{c\ell}$. 
In particular,
 \begin{align*}
    \mathcal{O}(1): & \lim_{(\bb,\cc)\rightarrow (0,0)} \frac{1}{1}  = 1 &\Rightarrow\hspace{0.5cm}&\textrm{Relevant} \\
    \mathcal{O}(\bb): & \lim_{(\bb,\cc)\rightarrow (0,0)} \frac{\bb}{\bb^2}  = \pm\infty &\Rightarrow\hspace{0.5cm}&\textrm{Relevant} \\
    \mathcal{O}(\cc): & \lim_{(\bb,\cc)\rightarrow (0,0)} \frac{\cc}{\bb\cc} = \pm\infty &\Rightarrow\hspace{0.5cm}&\textrm{Relevant} \\
    \mathcal{O}(\bb\cc): & \lim_{(\bb,\cc)\rightarrow (0,0)} \frac{\bb\cc}{\bb\cc} = 1 &\Rightarrow\hspace{0.5cm}&\textrm{Relevant} \\
    \mathcal{O}(\bb^2): & \lim_{(\bb,\cc)\rightarrow (0,0)} \frac{\bb^2}{\bb\cc} = \lim_{(\bb,\cc)\rightarrow (0,0)}\frac{\bb}{\cc} =   \textrm{indefinite}&\Rightarrow\hspace{0.5cm}&\textrm{Relevant} \\
    \mathcal{O}(\cc^2): & \lim_{(\bb,\cc)\rightarrow (0,0)} \frac{\cc^2}{\bb\cc} = \lim_{(\bb,\cc)\rightarrow (0,0)} \frac{\cc}{\bb} = \textrm{indefinite}&\Rightarrow\hspace{0.5cm}&\textrm{Relevant}  \\
    \mathcal{O}(\bb^2\cc): & 
    \lim_{(\bb,\cc)\rightarrow (0,0)}\frac{\bb^2\cc}{1} = \lim_{(\bb,\cc)\rightarrow (0,0)} \frac{\bb^2\cc}{\bb^2} = \lim_{(\bb,\cc)\rightarrow (0,0)}\frac{\bb^2\cc}{\bb\cc} = 0
    &\Rightarrow\hspace{0.5cm}& \textrm{Subdominant} 
   \end{align*}
Every other term of higher order will again be subdominant relative to the terms in $H_{c\ell}$. Therefore, the terms relevant for the classical asymptotic behavior are those proportional to the constant, $b,c,bc,b^2$ and $c^2$.

Let $M$ denote the number of values of $k$ summed over in the sum in equation \eqref{H most general} and in \eqref{H final}, which may be (countable) infinity, so that the label set summed over in these equations can, without loss of generality, be chosen as $\{1, \dots, M\}$.
Calling $\Lambda = \frac{(4\pi)^n}{2G\gamma^2}$ for simplicity, we obtain the following system of equations
enforcing the correct (na\"ive) classical limit:
\begin{align}\label{asymptotic behavior}
    \begin{split}
    \mathcal{O}(1): & \hspace{0.5cm} - \Lambda\gamma^2 |p_c|^{\frac{n-1}{2}} = a_0 +\sum_{k=1}^M a_k \\
    \mathcal{O}(\bb): & \hspace{0.5cm} 0 = \sum_{k=1}^M c_k A_k  \\
    \mathcal{O}(\cc): & \hspace{0.5cm} 0 = \sum_{k=1}^M b_k B_k \\
    \mathcal{O}(\bb\cc): & \hspace{0.3cm} -2 \Lambda|p_c|^{\frac{n+1}{2}} = \sum_{k=1}^M d_k A_k B_k \\
    \mathcal{O}(\bb^2): & \hspace{0.5cm} 2\Lambda|p_c|^{\frac{n-1}{2}} = \sum_{k=1}^M a_k A_k^2  \\
    \mathcal{O}(\cc^2): & \hspace{0.5cm} 0 = \sum_{k=1}^M a_k B_k^2 
    \end{split}
   \end{align}

\subsection{Choice of Lapse}\label{Choice of Lapse}
    
With the notion of classical limit made precise above, it is natural at this point to remark on why we have not included a certain common, and usually convenient, lapse choice in our derivations. Classically, this choice of lapse, which decouples the dynamics in $(b,p_b),(c,p_c)$ parts, is
\begin{align}\label{Ncl}
        N = \frac{\gamma}{b}\sgn{(p_c)}\sqrt{|p_c|},
\end{align}
resulting in
\begin{align}\label{Hcl[Ncl]}
    H_{c\ell}[N] = -\frac{1}{2G\gamma}\left(p_b\left(b + \frac{\gamma^2}{b}\right) + 2cp_c\right) = H_b[N_{c\ell}] + H_c[N_{c\ell}].
\end{align}
As convenient as this choice of lapse is, the presence of the $1/b$ factor complicates the definition of a corresponding operator on the Bohr Hilbert space.

Polymerized versions of (\ref{Hcl[Ncl]}) have indeed been introduced in recent works, yielding effective Hamiltonians that keep the decoupling property 
%and aids in making semiclassical predictions 
\cite{Corichi:2015xia,Ashtekar:2018cay,Ashtekar:2023cod}. 
In \cite{Corichi:2015xia}, for example, this is achieved by replacing $1/b$ with the function
\begin{align}
\label{litf}
f(b) = \frac{\delta_b}{\sin(\delta_b b)} 
\end{align}
where $\delta_b$ is a constant, $\mu$. (In \cite{Ashtekar:2018cay,Ashtekar:2023cod}, the same substitution is made, but with $\delta_b$ depending on both $b$ and $p_b$ --- see section \ref{aossect}.) As required, this is asymptotically equal to $1/b$ in the classical limit
$b \rightarrow 0$. 
Though \cite{Corichi:2015xia} introduces an operator on the Bohr Hilbert space, it is only for the bare Hamiltonian constraint \textit{without} lapse (that is, for lapse equal to $1$) --- polymerization of the lapse is only presented in the effective Hamiltonian. In \cite{Ashtekar:2018cay}, in the note at the end of its Appendix A, a strategy is suggested for constructing a fully quantum Hamiltonian operator, but is not carried out. 
%As far as we are aware, no definition of an operator corresponding to Hamiltonian constraint with lapse \eqref{Ncl} has appeared in the literature before the present work. 

Nevertheless, as a multiplicative operator, one can in fact show 
\eqref{litf} is densely defined on $\mathcal{H}_{\text{Bohr}}$. For example, if we choose the domain
$\mathcal{D}:=\sin(\mu b) \mathcal{H}_{\text{Bohr}}$, one can check that 
\begin{equation}
\lim_{\epsilon\rightarrow 0} \left|\left| e^{i\mu b} 
- \frac{e^{i\mu b}\sin(\mu b)}{i\epsilon + \sin(\mu b)}\right|\right|_{\text{Bohr}} = 0,
\end{equation}
so that each element $e^{i\mu b}$ of the momentum basis of 
$\mathcal{H}_{\text{Bohr}}$ is the limit of a corresponding family
$\phi_\mu^\epsilon(b) := \frac{e^{i\mu b}\sin(\mu b)}{i\epsilon + \sin(\mu b)}$ in $\mathcal{D}$,
showing $\mathcal{D}$ dense in $\mathcal{H}_{\text{Bohr}}$.

That being said, central to analyses of loop quantizations of symmetry reduced models is the momentum representation, in which every operator takes the form of a countable linear combination of shift operators with possibly non-constant coefficients. 
To cast the multiplicative operator $f(b)$ in this form requires its Fourier decomposition, which, since it is periodic, is a series. 
Since, over a period, $f(b)$ is not square integrable and thus not absolutely integrable, its Fourier series decomposition exists only in the distributional sense, with an infinite number of non-zero terms. Explicitly, if we interpret it as the distribution defined by its Cauchy principal value, since it is odd, its Fourier series decomposition includes only sine terms, 
with coefficients given by
\begin{align}
b_n := \frac{2 \mu^2}{\pi} \int_0^{\pi/\mu} 
\frac{\sin(n\mu b)}{\sin(\mu b)} db = \left\{ \begin{array}{cc} 2\mu & \text{ for }n\text{ odd} \\ 0 & \text{ for }n\text{ even}\end{array}\right.,
\end{align}
yielding the Fourier series
\begin{align}
\nonumber
\sum_{n=1}^\infty b_n \sin(n \mu b) &= 2\mu\lim_{M\rightarrow \infty}\sum_{m=1}^M \sin((2m+1)\mu b) \\
&=  2 f(b) \lim_{M\rightarrow \infty}  \sin^2((M+1)\mu b)
\end{align}
which converges to $f(b)$ in the distributional sense. 

Note, however, that one could also quantize $1/b$ as \textit{any} periodic function asymptotic to $1/b$ as $b \rightarrow 0$. 
Any such function will again not be square integrable over a period and so will possess an infinite number of terms in its Fourier expansion. 
Furthermore, for a large class of such functions, 
an argument similar to that above can be used to show it is densely defined on $\mathcal{H}_{\text{Bohr}}$ 
--- for
example, if the function's absolute value is bounded 
by $\left|\frac{A}{\sin(\mu b)}\right|$ for some $A$ and $\mu$, the argument follows
from the above argument for $\frac{\mu}{\sin(\mu b)}$.
Thus, there is actually an infinite dimensional ambiguity in how to quantize $1/b$ on the Bohr Hilbert space:
$\hat{\frac{1}{b}} = \frac{\mu}{\sin \mu b}$, is not the only possible one.
To avoid infinite dimensional ambiguities such as this, 
we choose to restrict consideration to Hamiltonian operators with only a finite number of shift operators, thereby excluding lapses with dependence on $1/b$. In the following section, we will consider the even more restrictive requirement that the number of terms in Hamiltonian operator be \textit{minimal}.

\section{Minimality}\label{Minimality}

Following the motivation of \cite{Engle:2019zfp}, we consider a further requirement: that the Hamiltonian have a minimum number of terms, i.e., a minimum number of shift exponentials consistent with the other requirements imposed. We achieve this by finding the solution of (\ref{asymptotic behavior}) for which the maximal number of coefficients can be set equal zero. From $\mathcal{O}(\bb)$ and $\mathcal{O}(\cc)$, this immediately implies that $b_k=c_k=0$, reducing the system of equations to
\begin{align}\label{minimal system equations}
    \begin{split}
    \mathcal{O}(1): & \hspace{0.5cm} - \Lambda\gamma^2 |p_c|^{\frac{n-1}{2}} = a_0 +\sum_{k=1}^M a_k \\
    \mathcal{O}(\bb\cc): & \hspace{0.3cm} -2 \Lambda|p_c|^{\frac{n+1}{2}} = \sum_{k=1}^M d_k A_k B_k \\
    \mathcal{O}(\bb^2): & \hspace{0.5cm} 2\Lambda|p_c|^{\frac{n-1}{2}} = \sum_{k=1}^M a_k A_k^2  \\
    \mathcal{O}(\cc^2): & \hspace{0.5cm} 0 = \sum_{k=1}^M a_k B_k^2 
    \end{split}
   \end{align}
The case $M=1$ is trivially ruled out, since the last equation would require $a_1=0$ or $B_1 = 0$, which would be inconsistent with the other equations. 
Choosing $M=2$, we first look at $\mathcal{O}(c^2):$
\begin{align*}
%\label{Oc2}
    a_1 B_1^2 + a_2 B_2^2 = 0.
\end{align*}
 Since we cannot have both $a_1, a_2$ equal to zero (because of the $\mathcal{O}(\bb^2)$ equation), then if we set $a_1 = 0$ we automatically must have $B_2 = 0$. 
 $d_2$ then appears nowhere in the remaining equations, 
 so that minimality forces $d_2=0$. 
 The solution for the remaining parameters is then
\begin{align*}
    a_2 = \frac{2\Lambda |p_c|^\frac{n-1}{2}}{A_2^2}, \hspace{0.5cm} d_1 = -\frac{2\Lambda|p_c|^\frac{n+1}{2}}{A_1B_1}, \hspace{0.5cm} a_o = -\Lambda|p_c|^\frac{n-1}{2}\left(\gamma^2 + \frac{2}{A_2^2}\right)
\end{align*}
for $A_1,A_2,B_1$ real functions of $|p_c|$
non-vanishing for $|p_c|\neq 0$, but otherwise arbitrary.

The minimal Hamiltonian can then be written as
\begin{align}\label{minimal Hamiltonian}
    H &= -\Lambda |p_b|^{n+1}|p_c|^\frac{n-1}{2}\sgn (p_bp_c)\left(\left(\gamma^2 + \frac{2}{A_2^2}\right) + 2|p_c|\sgn (p_bp_c)\frac{\sin(A_1b)}{A_1}\frac{\sin\left(B_1\frac{c}{|p_b|}\right)}{B_1} - 2\frac{\cos(A_2b)}{A_2^2}   \right) \nonumber \\
    &= -\frac{V^{n+1}\sgn(b)}{8\pi G\gamma^2p_c} \left(\gamma^2 + 2p_c\sgn p_b\frac{\sin(A_1b)}{A_1}\frac{\sin\left(B_1\frac{c}{|p_b|}\right)}{B_1} + \frac{4\sin^2\left(\frac{A_2}{2}b\right)}{A_2^2}\right),
\end{align}
and the full Hamiltonian operator, using \eqref{operator Frankstein}, becomes
\begin{align}
\nonumber
8\pi G \gamma^2 \hat{H} = &
\frac{\hat{V}^{(n+1)}\sgn(p_b)}{-\hat p_c} \left(\gamma^2+\frac{2}{A_2^2}\right)\\
\nonumber
& 
+e^{\frac{iA_1 b}{2}}e^{\frac{iB_1 c}{2|p_b|}}
\frac{\hat{V}^{(n+1)}}{2A_1 B_1}e^{\frac{iB_1 c}{2|p_b|}}e^{\frac{iA_1 b}{2}} 
-e^{\frac{iA_1 b}{2}}e^{\frac{-iB_1 c}{2|p_b|}}
\frac{\hat{V}^{(n+1)}}{2A_1 B_1}e^{\frac{-iB_1 c}{2|p_b|}}e^{\frac{iA_1 b}{2}} \\
\nonumber
& 
-e^{\frac{-iA_1 b}{2}}e^{\frac{iB_1 c}{2|p_b|}}
\frac{\hat{V}^{(n+1)}}{2A_1 B_1}e^{\frac{iB_1 c}{2|p_b|}}e^{\frac{-iA_1 b}{2}} 
+ e^{\frac{-iA_1 b}{2}}e^{\frac{-iB_1 c}{2|p_b|}}
\frac{\hat{V}^{(n+1)}}{2A_1 B_1}e^{\frac{-iB_1 c}{2|p_b|}}e^{\frac{-iA_1 b}{2}} \\
&
+e^{\frac{iA_2 b}{2}}
\frac{\hat{V}^{(n+1)}\sgn(p_b)}{A_2^2 p_c} e^{\frac{iA_2 b}{2}}
+e^{\frac{-iA_2 b}{2}}
\frac{\hat{V}^{(n+1)}\sgn(p_b)}{A_2^2 p_c} e^{\frac{-iA_2 b}{2}}.
\end{align}

\section{Comparison with prescriptions in the literature}
\label{Comparison Literature}

Most proposals for a Hamiltonian in loop quantum Kantowski-Sachs are only for an effective Hamiltonian from which physical predictions can be made \cite{Campiglia:2007pb,Chiou:2008eg,Joe:2014tca, Ashtekar:2018cay, Bodendorfer:2019cyv, Assanioussi:2019twp, Sartini:2020ycs}, while a few others seek to build a full Hamiltonian operator \cite{Modesto:2005zm,Ashtekar:2005qt,Corichi:2015xia, Sartini:2020ycs}. 
While comparing full Hamiltonian operators may be harder, due to possible differences in the operators ordering, it is easy to verify whether a proposed effective Hamiltonian matches one of the selected 
solutions \eqref{H final} (and whether it is minimal). 

What distinguishes each approach to KS is the prescription for quantizing the curvature and connection factors in the Hamiltonian constraint ---
specifically how the curves used to regularize these factors are determined by the edges, coordinates, or metric of the fiducial cell in terms of the smallest non-zero area eigenvalue $\Delta$.
The coordinate lengths of the components of these curves is what determines the coefficients of the connection components in the exponentials appearing in the final effective Hamiltonian constraint.
From the form \eqref{H preserved Bohr}, we see that diffeomorphism covariance forces some of these coefficients to be non-constant --- specifically, the coefficient of $c$ must depend inversely on $p_b$, a fact which can be seen more directly from the flows \eqref{resulting flows}. $\mu_0$-schemes \cite{Modesto:2005zm,Ashtekar:2005qt,Campiglia:2007pb,Corichi:2015xia, Assanioussi:2019twp}, for which all such coordinate edge lengths are constant, are thus excluded by covariance. 
%
% N.B.: I commented out the sentence below because some authors
% think the fiducial cell is actually physical, so saying this
% will only muddy the argument. The key point is covariance under
% active residual diffeos. 
%
%%This is consistent with the fact that the results of these proposals are not invariant under rescaling of the fiducial cell.  
%
Instead, covariance points towards some sort of $\bar{\mu}$-scheme, as in \cite{Chiou:2008eg,Joe:2014tca}. In the first two subsections below, we specify the relation of such proposals to our results. In last two subsections, we then discuss other proposals in the literature with non-constant coordinate edge lengths.

\subsection{$n = 0:$ Proper time case}

If we take $N = 1$ --- which is equivalent to $n = 0$ in the lapse (\ref{lapse volume}) --- $A_1 = \sqrt{\frac{\Delta}{|p_c|}}$, $A_2 = 2A_1$, and $B_1 = \sqrt{|p_c|\Delta}$, we obtain
\begin{align}
    \label{Joe/Singh}  H = -\frac{|p_b|\sqrt{|p_c|}}{2G\gamma^2\Delta}\left(\frac{\gamma^2\Delta}{|p_c|} + 2\sin\left(\sqrt{\frac{\Delta}{|p_c|}}b\right)\sin\left(\sqrt{|p_c|\Delta}\frac{c}{|p_b|}\right) + \sin^2\left(\sqrt{\frac{\Delta}{|p_c|}}b\right)\right),
\end{align}
which matches the results obtained by Joe and Singh in \cite{Joe:2014tca}, and Cortez, Cuervo, Morales-Técotl and Ruelas in \cite{Cortez:2017alh}, for $p_b,p_c>0$.  

\subsection{$n = 1:$ Harmonic time gauge}
For $n = 1$, with the same assumptions and restrictions as above, we find 
\begin{align}
    \label{Chiou}  H = -\frac{2\pi p_b^2p_c}{G\gamma^2\Delta}\left(\frac{\gamma^2\Delta}{p_c} + 2\sin\left(\sqrt{\frac{\Delta}{p_c}}b\right)\sin\left(\sqrt{p_c\Delta}\frac{c}{p_b}\right) + \sin^2\left(\sqrt{\frac{\Delta}{p_c}}b\right)\right),
\end{align}
matching the result obtained by Chiou in \cite{Chiou:2008eg}.

\subsection{AOS Prescription}
\label{aossect}

Different from the above cases is that recently introduced by Ashtekar, Olmedo, and Singh (AOS) \cite{Ashtekar:2018cay,Ashtekar:2023cod}.
As usual, when constructing the Hamiltonian for loop quantum Kantowski-Sachs, they begin by regularizing the curvature in terms of parallel transports around finite loops, with edges in the $x$-direction with coordinate lengths $L_o \delta_c$ and edges within $x=$constant surfaces along geodesics of the fiducial unit sphere metric $d\Omega^2 = d\theta^2 + \sin^2 \theta d\phi^2$ with (dimensionless) `lengths' $2\pi \delta_b$ relative to $d\Omega^2$.
The key requirements in this model are that
\begin{enumerate}
    \item (as in \cite{Corichi:2015xia,Olmedo:2017lvt}) $\delta_b$ and $\delta_c$ be Dirac observables --- i.e., are constant on dynamical trajectories --- and 
    \item %(unlike in \cite{Corichi:2015xia,Olmedo:2017lvt}) 
    \textit{at the transition surface that replaces the classical singularity},
    the regularizing loops enclose a physical area equal the area gap $\Delta$ when the Hamiltonian constraint is satisfied. 
\end{enumerate} 
In the resulting effective model, the expansion and shear diverge at the horizon just as in classical general relativity, so that the model matches general relativity in this regime, exactly as it should. This is a major advantage of AOS over the $\overline{\mu}$-schemes introduced so far, an advantage shared by \cite{Corichi:2015xia,Olmedo:2017lvt}, suggesting that it is the first of the above requirements that ensures this. 
In contrast to \cite{Corichi:2015xia,Olmedo:2017lvt}, the AOS model further ensures, as in the $\overline{\mu}$-schemes, that the transition surface always occurs in a regime where quantum gravity effects are expected to be relevant, namely when the Kretschmann scalar is on the order of the Planck scale. 
In addition to these advantages which no other model simultaneously shares, the authors of \cite{Ashtekar:2018cay,Ashtekar:2023cod} have extended their analysis to the exterior of the black hole and explored the global structure of the resulting maximally extended effective space-time. It is thus the most well-developed and physically viable model proposed so far in the literature.

The conditions 1. and 2. still leave considerable freedom in the definitions of $\delta_b$ and $\delta_c$, and there is also the freedom in the choice of lapse. The authors choose to use these freedoms in order to decouple the dynamics of the $(b,p_b)$ and $(c,p_c)$ degrees of freedom, allowing for exact analytic solutions to the effective equations. Specifically, the lapse \eqref{Ncl} is the choice made by AOS.
With this choice, the regularized effective Hamiltonian constraint becomes
\begin{align}
\label{AOSHam}
H = -\frac{1}{2G\gamma}\left[\left(\frac{\sin(\delta_b b)}{\delta_b} +\frac{\gamma^2 \delta_b}{\sin(\delta_b b)}\right)p_b
+2\frac{\sin(\delta_c c)}{c}p_c\right] .
\end{align}
In order for this effective Hamiltonian to retain the decoupling of the $b$ and $c$ degrees of freedom in the classical theory, they further require that
\begin{align}
\label{separate deltas}
\delta_b\text{ depend only on }(b,p_b)\text{ and }\delta_c\text{ only on }(c,p_c).
\end{align}
As convenient as it is to have an exactly soluble model, the choice of lapse and the conditions \eqref{separate deltas}, respectively, come at the cost of (a.) the corresponding Hamiltonian constraint \textit{operator} having an infinite number of shift terms if implemented on the usual Bohr Hilbert space motivated from loop quantum gravity, and (b.) the effective Hamiltonian constraint \textit{not being covariant under residual diffeomorphisms}.
That the corresponding operator on the Bohr Hilbert space 
must have an infinite number of shift operators follows from our discussion in section \ref{Choice of Lapse}.

To see (b.), we must be more explicit. Concretely, in AOS,
in the large mass limit, one can understand $\delta_b(b,p_b)$ to be obtained as the solution of the transcendental system of two equations consisting in the first equation in each of (2.12) and (2.13) in \cite{Ashtekar:2023cod}, and $\delta_c(c,p_c)$ is obtained as the solution of the system consisting in the second equation in each of these. By using (2.13) to eliminate $m_b$ in (2.12), one sees that not only \textit{can} $\delta_b$ depend only on $b$ and $p_b$, but it must depend on both non-trivially. Likewise $\delta_c$ must depend on both $c$ and $p_c$ non-trivially.
As a consequence, under the flow \eqref{resulting flows},
the argument of the first sine in \eqref{AOSHam} is non-constant,
$\dot{\left(\delta_b b\right)}= \frac{\partial \delta_b}{p_b}p_b b \neq 0$, 
forcing the AOS effective Hamiltonian to be not covariant under residual diffeomorphisms. 

Note that, even without the specific prescription of AOS for fixing $\delta_b(b,p_b)$ and $\delta_c(c,p_c)$, the assumption \eqref{separate deltas} alone is enough to force incompatibility with the form \eqref{classical analogue} we have shown is required by residual diffeomorphism covariance and preservation of the Bohr Hilbert space, in which $\delta_c$ must depend on $p_b$ with a very specific dependence.
This suggests that the desire to maintain decoupling of the $b$ and $c$ degrees of freedom in the effective theory, as convenient as it is, is incompatible with simultaneous residual diffeomorphism covariance and the existence of a corresponding operator preserving the Bohr Hilbert space. That is: if one desires both of the latter two properties, then the $b$ and $c$ degrees of freedom are forced to interact. 

It must be emphasized that, even though the precise effective Hamiltonian constraint of AOS is not covariant under residual diffeomorphisms, the \textit{key physical predictions} calculated so far, such as the universal upper bound on all scalar curvatures, \textit{are} invariant under residual diffeomorphisms. 
Furthermore, the works \cite{Ashtekar:2023cod,Ashtekar:2018cay} never suggest that their proposed effective Hamiltonian have a corresponding quantum operator on the Bohr Hilbert space. Indeed, they explicitly mention the construction of a corresponding operator and associated Hilbert space as an open problem at the end of appendix A in \cite{Ashtekar:2018cay}, and provide a strategy for constructing an alternative quantum framework.

However, diffeomorphism symmetry being the basic symmetry of gravity, there is good motivation
%epistemic motivation 
to seek an effective Hamiltonian that is \textit{exactly} covariant under residual diffeomorphisms. Because of this, it might be valuable to attempt a modification of AOS in which the condition \eqref{separate deltas}, and hence the decoupling of the two degrees of freedom, is dropped, and in which such exact covariance is imposed in its place. Such a model would be mathematically more complex, but, potentially, physically more compelling, including all of the physically compelling features of AOS as well as the exact residual diffeomorphism covariance of the $\overline{\mu}$ models.

\subsection{Newer proposals}

Some newer proposals, with different approaches, are worth mentioning:
\begin{itemize}
    \item Assanioussi and Mickel \cite{Assanioussi:2020ezr} proposed an effective Hamiltonian constructed via regularized Thiemann identities, in the $\bar \mu$ scheme. Their starting point differs from ours --- the Hamiltonian from the full theory, with a Euclidian and a Lorentzian component, while our approach uses the symmetry reduced Hamiltonian \eqref{Hcl}, in which these two terms are not distinguished  --- so the final result is expected to be different. However, their result does lie in the family \eqref{H final} selected by residual diffeomorphism covariance and discrete symmetries, and our minimal result has the same form as the Euclidian part calculated by them.
    % , and the results would be identical in a formulation where $\gamma = i$.   
    
    \item Bodendorfer, Mele and Munch \cite{Bodendorfer:2019nvy} introduce new pairs of canonical variables,
    \begin{align*}
        v_k := \frac{\gamma p_b|p_c|}{2^\frac{14}{3} b}, \hspace{0.5cm} v_j := \frac{p_b}{8b}(cp_c-bp_b), \hspace{0.5cm} k := \frac{2^\frac{11}{3}bc}{\gamma^2p_b\sgn p_c}, \hspace{0.5cm} j := \frac{4b}{\gamma p_b},
    \end{align*}
    in order to have a relation $\mathcal{K} \propto k^2$ for the Kretschmann scalar,  inspired by the relation $R \propto b^2$ that appears using $(b,v)$ variables in the homogeneous isotropic case. They use the lapse \eqref{Ncl}, and the effective Hamiltonian density is obtained by a polymerization of the variables $k$ and $j$, resulting in 
    \begin{align*}
        \mathcal{H}_{eff} = 3v_k\frac{\sin(\lambda_kk)}{\lambda_k}\frac{\sin(\lambda_jj)}{\lambda_j} + v_j\frac{\sin^2(\lambda_jj)}{\lambda_j}.
    \end{align*}
    
    However, both $v_j,v_k$ are proportional to $1/b$, which requires an infinite number of terms to be represented in the Bohr Hilbert space, as discussed in Section \ref{Choice of Lapse}.  
    Moreover, in order to ensure covariance of the effective Hamiltonian under rescaling of the fiducial cell by a factor $\alpha$, the parameter $\lambda_j$ --- a constant on phase space --- is defined to scale as $\lambda_j \mapsto \alpha \lambda_j$.  While such a definition is possible to ensure covariance under passive residual diffeomorphisms, there is no such freedom for active diffeomorphisms --- arising from a flow on the phase space --- where constants are simply constant. As a consequence, their effective Hamiltonian is not covariant under active residual diffeomorphisms, explaining why it does not fall in to the class \eqref{H final} that we have selected above. Also, $k$ is quadratic in components of the connections, so the first term in their Hamiltonian could not come from parallel transports of the Ashtekar-Barbero connections. The fact that components of the Ashtekar-Barbero connection appear quadratically in one of the sines furthermore means that
    the Fourier transform of their effective Hamiltonian with respect to $b$ and $c$ must have uncountable support, impeding a corresponding operator from being densely defined on the usual Bohr Hilbert space reviewed in section \ref{Quantum Kinematics}.
    
    \item Sartini and Geiller \cite{Sartini:2020ycs} consider KS with cosmological constant incorporated via the unimodular formulation of gravity \cite{Unruh:1988in}, the main motivation being to solve the problem of time without introducing scalar matter. They propose the change of variables
    \begin{align*}
        p_1 := -\frac{c}{2\gamma},\hspace{0.5cm} v_1 := p_c,\hspace{0.5cm} p_2 := \frac{4b}{\gamma p_b}, \hspace{0.5cm} v_2 := - \frac{p_b^2}{8} .
    \end{align*}
    For their effective theory, they again choose the lapse \eqref{Ncl}, and polymerize $p_1$ and $p_2$, resulting in 
    \begin{align}
        H = 2\frac{\sin(\lambda_1 p_1)}{\lambda_1} v_1 
        + \frac{\sin(\lambda_2 p_2)}{\lambda_2} v_2
        - 2(1-\Lambda v_1)\frac{\lambda_2}{\sin(\lambda_2 p_2)}.
    \end{align}
    with $\lambda_i$ constants on phase space.
    The case here is similar to the one above, where their definition of how the constants $\lambda_i$ should rescale under changes of the fiducial cell make the effective Hamiltonian covariant under passive but not active diffeomorphisms. 

    The use of the classical lapse \eqref{Ncl} for the effective theory means that, if the effective Hamiltonian would arise from a quantum operator, then again the discussion of section \ref{Choice of Lapse} would apply. However, when proposing a quantum Hamiltonian \textit{operator}, the authors make use of a different lapse, the one corresponding the use of a unimodular clock, matching \eqref{lapse volume} for $n=-1$. 
    The Hilbert space on which the non-cosmological constant part of their operator acts is the usual Bohr Hilbert space.
    That being said, the polymerization of the connection variables in their operator remains the same as in their effective theory, and so again is not covariant under active residual diffeomorphisms, so that the non-cosmological constant part of the operator is not in the family \eqref{H final} we have selected. 
         
\end{itemize}

\section{Conclusion}

In this work, we were able to derive a family of Hamiltonian operators for the loop quantum Kantowski-Sachs framework, by imposing the quantum analogue of covariance under residual diffeomorphisms, as well as other physical criteria. In doing this, we avoid choosing a specific quantization prescription a priori. 

We further demonstrated that, for each choice of lapse, the requirement of minimality, that is, a minimal number of shift operators in the Hamiltonian constraint operator ---
a form of Occam's razor --- 
leads to a family of models parameterized by three functions of $p_c$.
For specific values of these parameters, 
the model matches proposals in the literature constructed by traditional quantization methods, specifically the $\overline{\mu}$-prescriptions obtained by Joe and Singh \cite{Joe:2014tca}, and Chiou \cite{Chiou:2008eg}. We emphasize, however, that the minimality principle is trustworthy only inasmuch as the other conditions imposed are complete --- in particular, we impose no condition relating the model's dynamics to a choice of full theory dynamics, a condition whose incorporation would likely force a non-minimal choice as defined here. 

We have also remarked on the relation of our work to other models 
in the literature, with particular attention to that of
of Ashtekar, Olmedo, and Singh (AOS) \cite{Ashtekar:2018cay,Ashtekar:2023cod}.
First, and most importantly, AOS, as well as \cite{Corichi:2015xia,Olmedo:2017lvt}, improves upon all previous works in that the classical limit is correctly imposed at the horizon --- a regime where (for macroscopic black holes) curvature is low compared to the Planck scale, so that no significant deviation from classical general relativity is expected. That prior models, including those using $\overline{\mu}$-schemes, failed to do this highlights that the condition for the correct classical limit imposed in the literature up until now, also used in the present paper, is not sufficient. Specifically, the usual condition that the arguments of complex exponentials, or, equivalently, of the sine functions, go to zero, motivated by a na\"ive $\ell_P \rightarrow 0$ limit, and equivalent to eigenvalues of extrinsic curvature going to zero --- is neither a necessary nor sufficient condition that \textit{space-time} curvature scalars go to zero.
%which is the space-time-diffeomorphism invariant condition for the classical regime. 

The AOS model additionally makes two choices to decouple the evolution of the two degrees of freedom of the model, rendering the dynamics exactly soluble: The choice of lapse, and the requirement that $\delta_b$ and $\delta_c$ depend, respectively, only on the $b$ and $c$ degrees of freedom.  As attractive as exact solubility is, the latter of these choices forces the effective Hamiltonian to not be exactly diffeomorphism covariant. The first of these choices forces the corresponding Hamiltonian constraint operator, if defined on the Bohr Hilbert space motivated by loop quantum gravity, to include an infinite number of shift operators. 

It must be emphasized that the key physical predictions of AOS \textit{are} covariant under residual diffeomorphisms. Nevertheless, we argue that exact diffeomorphism covariance of the full Hamiltonian is a compelling property, and that it should be possible to modify the AOS model to require such exact covariance if one gives up decoupling the two degrees of freedom of the model, while retaining all of the model's physically compelling features.
%which has made it the most well developed and physically compelling model up to this point in time.
One systematic path to find such a new model might be to use the program of the present work, but replacing the usual na\"ive classical limit used in section \ref{low curvature} with 
an appropriate corrected condition based on four dimensional curvature scalars. 
%We leave such exploration to the future efforts. 

\section*{Acknowledgements}

The authors thank Christopher Beetle for discussions helpful in obtaining the results presented in section \ref{Choice of Lapse}. 
This work was supported in part by NSF Grants PHY-1806290 and PHY-2110234.

% \bibliography{Article/References}

\begin{thebibliography}{10}

\bibitem{Rovelli:2004tv}
C.~Rovelli, {\em {Quantum gravity}}.
\newblock Cambridge Monographs on Mathematical Physics, Cambridge University
  Press, 2004.

\bibitem{Ashtekar:2004eh}
A.~Ashtekar and J.~Lewandowski, ``{Background independent quantum gravity: A
  Status report},'' {\em Class. Quant. Grav.}, vol.~21, p.~R53, 2004.

\bibitem{Thiemann:2007pyv}
T.~Thiemann, {\em {Modern Canonical Quantum General Relativity}}.
\newblock Cambridge Monographs on Mathematical Physics, Cambridge University
  Press, 2007.

\bibitem{Gambini:2011zz}
R.~Gambini and J.~Pullin, {\em {A first course in loop quantum gravity}}.
\newblock Oxford University Press, 2011.

\bibitem{Rovelli:2014ssa}
C.~Rovelli and F.~Vidotto, {\em {Covariant Loop Quantum Gravity}: {An
  Elementary Introduction to Quantum Gravity and Spinfoam Theory}}.
\newblock Cambridge Monographs on Mathematical Physics, Cambridge University
  Press, 2014.

\bibitem{Ashtekar:2017yom}
A.~Ashtekar and J.~Pullin, eds., {\em {Loop Quantum Gravity}: {The First 30
  Years}}, vol.~4 of {\em 100 Years of General Relativity}.
\newblock World Scientific, 2017.

\bibitem{Bojowald:2008zzb}
M.~Bojowald, ``{Loop quantum cosmology},'' {\em Living Rev. Rel.}, vol.~11,
  p.~4, 2008.

\bibitem{Ashtekar:2011ni}
A.~Ashtekar and P.~Singh, ``{Loop Quantum Cosmology: A Status Report},'' {\em
  Class. Quant. Grav.}, vol.~28, p.~213001, 2011.

\bibitem{Agullo:2016tjh}
I.~Agullo and P.~Singh, {\em {Loop Quantum Cosmology} {\rm in} {Loop Quantum
  Gravity}: {The First 30 Years}}, pp.~183--240.
\newblock World Scientific, 2017.

\bibitem{Lewandowski:2005jk}
J.~Lewandowski, A.~Okolow, H.~Sahlmann, and T.~Thiemann, ``{Uniqueness of
  diffeomorphism invariant states on holonomy-flux algebras},'' {\em Commun.
  Math. Phys.}, vol.~267, pp.~703--733, 2006.

\bibitem{Ashtekar:2012cm}
A.~Ashtekar and M.~Campiglia, ``{On the Uniqueness of Kinematics of Loop
  Quantum Cosmology},'' {\em Class. Quant. Grav.}, vol.~29, p.~242001, 2012.

\bibitem{Engle:2016hei}
J.~Engle and M.~Hanusch, ``{Kinematical uniqueness of homogeneous isotropic
  LQC},'' {\em Class. Quant. Grav.}, vol.~34, no.~1, p.~014001, 2017.

\bibitem{Engle:2016zac}
J.~Engle, M.~Hanusch, and T.~Thiemann, ``{Uniqueness of the Representation in
  Homogeneous Isotropic LQC},'' {\em Commun. Math. Phys.}, vol.~354, no.~1,
  pp.~231--246, 2017.
\newblock [Erratum: Commun.Math.Phys. 362, 759--760 (2018)].

\bibitem{Engle:2018zbe}
J.~Engle and I.~Vilensky, ``{Deriving loop quantum cosmology dynamics from
  diffeomorphism invariance},'' {\em Phys. Rev. D}, vol.~98, no.~2, p.~023505,
  2018.

\bibitem{Engle:2019zfp}
J.~Engle and I.~Vilensky, ``{Uniqueness of minimal loop quantum cosmology
  dynamics},'' {\em Phys. Rev. D}, vol.~100, no.~12, p.~121901, 2019.

\bibitem{Modesto:2005zm}
L.~Modesto, ``{Loop quantum black hole},'' {\em Class. Quant. Grav.}, vol.~23,
  pp.~5587--5602, 2006.

\bibitem{Ashtekar:2005qt}
A.~Ashtekar and M.~Bojowald, ``{Quantum geometry and the Schwarzschild
  singularity},'' {\em Class. Quant. Grav.}, vol.~23, pp.~391--411, 2006.

\bibitem{Campiglia:2007pb}
M.~Campiglia, R.~Gambini, and J.~Pullin, ``{Loop quantization of spherically
  symmetric midi-superspaces : The Interior problem},'' {\em AIP Conf. Proc.},
  vol.~977, no.~1, pp.~52--63, 2008.

\bibitem{Chiou:2008eg}
D.-W. Chiou, ``{Phenomenological dynamics of loop quantum cosmology in
  Kantowski-Sachs spacetime},'' {\em Phys. Rev. D}, vol.~78, p.~044019, 2008.

\bibitem{Joe:2014tca}
A.~Joe and P.~Singh, ``{Kantowski-Sachs spacetime in loop quantum cosmology:
  bounds on expansion and shear scalars and the viability of quantization
  prescriptions},'' {\em Class. Quant. Grav.}, vol.~32, no.~1, p.~015009, 2015.

\bibitem{Corichi:2015xia}
A.~Corichi and P.~Singh, ``{Loop quantization of the Schwarzschild interior
  revisited},'' {\em Class. Quant. Grav.}, vol.~33, no.~5, p.~055006, 2016.

\bibitem{Cortez:2017alh}
J.~Cortez, W.~Cuervo, H.~A. Morales-T\'ecotl, and J.~C. Ruelas, ``{Effective
  loop quantum geometry of Schwarzschild interior},'' {\em Phys. Rev. D},
  vol.~95, no.~6, p.~064041, 2017.

\bibitem{Ashtekar:2018cay}
A.~Ashtekar, J.~Olmedo, and P.~Singh, ``{Quantum extension of the Kruskal
  spacetime},'' {\em Phys. Rev. D}, vol.~98, no.~12, p.~126003, 2018.

\bibitem{Bodendorfer:2019cyv}
N.~Bodendorfer, F.~M. Mele, and J.~M\"unch, ``{Effective Quantum Extended
  Spacetime of Polymer Schwarzschild Black Hole},'' {\em Class. Quant. Grav.},
  vol.~36, no.~19, p.~195015, 2019.

\bibitem{Sartini:2020ycs}
F.~Sartini and M.~Geiller, ``{Quantum dynamics of the black hole interior in
  loop quantum cosmology},'' {\em Phys. Rev. D}, vol.~103, no.~6, p.~066014,
  2021.

\bibitem{Ashtekar:2023cod}
A.~Ashtekar, J.~Olmedo, and P.~Singh, {\em Regular Black Holes from Loop
  Quantum Gravity {\rm in} Regular Black Holes: Towards a New Paradigm of
  Gravitational Collapse}, pp.~235--282.
\newblock Springer Nature Singapore, 2023.

\bibitem{Olmedo:2017lvt}
J.~Olmedo, S.~Saini, and P.~Singh, ``{From black holes to white holes: a
  quantum gravitational, symmetric bounce},'' {\em Class. Quant. Grav.},
  vol.~34, no.~22, p.~225011, 2017.

\bibitem{Bodendorfer:2019nvy}
N.~Bodendorfer, F.~M. Mele, and J.~M\"unch, ``{(b,v)-type variables for black
  to white hole transitions in effective loop quantum gravity},'' {\em Phys.
  Lett. B}, vol.~819, p.~136390, 2021.

\bibitem{Assanioussi:2020ezr}
M.~Assanioussi and L.~Mickel, ``{Loop effective model for Schwarzschild black
  hole interior: a modified $\bar \mu$ dynamics},'' {\em Phys. Rev. D},
  vol.~103, no.~12, p.~124008, 2021.

\bibitem{Assanioussi:2019twp}
M.~Assanioussi, A.~Dapor, and K.~Liegener, ``{Perspectives on the dynamics in a
  loop quantum gravity effective description of black hole interiors},'' {\em
  Phys. Rev. D}, vol.~101, no.~2, p.~026002, 2020.

\bibitem{Unruh:1988in}
W.~G. Unruh, ``{A Unimodular Theory of Canonical Quantum Gravity},'' {\em Phys.
  Rev. D}, vol.~40, p.~1048, 1989.

\end{thebibliography}
% \bibliographystyle{ieeetr}
%% Reason for ieeetr: (1.) bibliography is in order of citation,
%% (2.) titles are given for articles, but (3.) no full author first names
%% are given, only initials, and (4.) format for authors is same for
%% both articles and books
%%

\end{document}